\documentclass[useAMS,usenatbib,usegraphicx,fleqn]{mn2e}
\usepackage{graphicx}

\usepackage{subfigure}
\usepackage{amssymb}
\usepackage{epstopdf}
\usepackage{rotating}
\usepackage{amsmath}
\usepackage{color}
\usepackage{natbib}
\usepackage[usenames,dvipsnames]{xcolor}
\usepackage{pdflscape}

\title[Co-evolution of BCGs and ICL using CLASH]{Co-evolution of Brightest Cluster Galaxies and Intracluster Light using CLASH}
\author[Burke, Hilton  \&  Collins]{Claire Burke$^{1}$\thanks{E-mail: burke.astro@gmail.com (C.B.)}, Matt Hilton$^{1}$  \&  Chris Collins$^{2}$\\
\vspace{0.4cm}\\
\parbox{\textwidth}{\raggedright 
$^{1}$~Astrophysics \& Cosmology Research Unit, School of Mathematics, Statistics \& Computer Science, University of KwaZulu-Natal, 
Westville Campus, Durban 4000, South Africa.\\
$^{2}$~Astrophysics Research Institute, Liverpool John Moores University, IC2, Liverpool Science Park, 146 Brownlow Hill, Liverpool, L3~5RF, UK.\\
}
}
\begin{document}

\date{Accepted 2015 February 27. Received 2015 January 31; in original form 2014 September 23 }

\pagerange{\pageref{firstpage}--\pageref{lastpage}} \pubyear{2011}

\maketitle

\label{firstpage}

\begin{abstract}
We examine the stellar mass assembly in galaxy cluster cores using data from the Cluster Lensing and Supernova survey with Hubble (CLASH). We measure the growth of brightest cluster galaxy (BCG) stellar mass, the fraction of the total cluster light which is in the intracluster light (ICL) and the numbers of mergers that occur in the BCG over the redshift range of the sample, 0.18$<z<0.90$. We find that BCGs grow in stellar mass by a factor of 1.4 on average from accretion of their companions, and this growth is reduced to a factor of 1.2 assuming 50\% of the accreted stellar mass becomes ICL, in line with the predictions of simulations. We find that the ICL shows significant growth over this same redshift range, growing by a factor of of 4--5 in its contribution to the total cluster light. This result is in line with our previous findings for ICL at higher redshifts,  however our measured growth is somewhat steeper than is predicted by simulations of ICL assembly. We find high mass companions and hence major merging (mergers with objects of masses $\geq$1/2 of the BCG) to be very rare for our sample. We conclude that minor mergers (mergers with objects with masses $<$ 1/2 of the BCG) are the dominant process for stellar mass assembly at low redshifts, with the majority of the stellar mass from interactions ending up contributing to the ICL rather than building up the BCG. From a rough estimate of the stellar mass growth of the ICL we also conclude that the majority of the ICL stars must come from galaxies which fall from outside of the core of the cluster, as predicted by simulations. It appears that the growth of the ICL is the major evolution event in galaxy cluster cores during the second half of the lifetime of the Universe.

\end{abstract}


\section{Introduction}\label{intro}
Having formed from the largest density perturbations remaining after the Big Bang, galaxy clusters give us insight into the conditions in the early Universe and can provide constraints for cosmology. 

The most massive galaxies in the Universe reside in the cores of galaxy clusters. Brightest cluster galaxies (BCGs) generally sit close to or at the centre of the gravitational potential of their host cluster. BCGs are thought to have formed and evolved by a hierarchical merging process, which would cause them to grow gradually over time. This gradual assembly is predicted by cosmological simulations of structure formation \citep[e.g.][]{DeLucia_07}, however observations of the stellar masses of BCGs show that they have assembled the majority of their mass fairly early on, with some studies finding as much as 95\% of present day BCG masses being already in place by redshift $z\sim1$ \citep[equivalent to about half-way back to the Big Bang;][] {Collins_09, Stott_10}. Subsequently a range of mass growth estimates have been reported over the timescale between $0<z<1$, with some studies showing more growth over this redshift range \citep[e.g.][]{Lidman_12, Lin_13} necessitating a less extreme and rapid early assembly. In any case the observed masses of BCGs at high redshifts are generally larger than are predicted by simulations. This apparent rapid and early assembly of massive galaxies in cluster cores is a major challenge for models of galaxy formation and remains a topic of much debate.

Galaxy cluster cores are observed to be pervaded by a diffuse, low brightness, stellar component known as intracluster light (ICL). The ICL is made up of stars which are thought to be gravitationally bound to the cluster potential rather than any specific galaxy. The origin of the ICL is still unclear at the present time, although it is generally thought that it can be produced by interactions between galaxies in clusters, such as tidal stripping and merging. In nearby clusters the ICL, whilst spread between the majority of the cluster galaxies, is often found to be centrally concentrated around the BCG \citep[e.g.][]{Mihos_05, Rudick_11}. As such it seems that the evolution and assembly histories of BCGs and the ICL are probably intertwined. Observations and simulations of the ICL show a large growth between $0<z<1$ \citep[e.g.][]{Krick_07, Murante_07, Rudick_11, B12, Contini_14} and when combined with the expected numbers of mergers in cluster cores and onto BCGs \citep[e.g.][]{BC13, Edwards_12, Laporte_13, Lidman_13} this presents a scenario of the ICL being built up by interactions of galaxies with the BCG. This mechanism for ICL growth would allow BCG masses to remain relatively unchanged, or to only grow by a small amount over this redshift range if the majority of the stellar mass from mergers ends up in the ICL rather than centrally on the BCG. Simulations of ICL buildup and mergers in cluster cores show that a large fraction of merging stellar mass (between 30--80\%, see e.g. \citealt{Conroy_07, Puchwein_10, Laporte_13}) will end up residing in the diffuse halo of the BCG or ICL, however this scenario has not yet been observationally confirmed.

In this paper we aim to examine the growth and buildup of BCGs and the ICL, and test the scenario of ICL growth through interactions between galaxies in the cluster core. In order to do this we use the X-ray selected Cluster Lensing and Supernova survey with Hubble \citep[CLASH;][]{Postman_12} cluster sample which covers the redshift range $0.18<z<0.89$. We measure the BCG stellar masses, the contributions of the ICL and BCG to the total cluster star light, and the number of mergers that the BCG may have experienced over this redshift range. Our approach represents a self-consistent analysis of the mass assembly in galaxy clusters including BCG assembly, ICL and BCG companion galaxies.

The structure of this paper is as follows: Section~\ref{data_s} contains the details of the sample studied; in Section~\ref{analysis_section} the methods and corrections used are described; in Section~\ref{results_s} we present the results of our measurements; the results are discussed in Section~\ref{discussion} and conclusions drawn in Section~\ref{conc_s}.
Throughout this paper we adopt a $\Lambda$CDM cosmology with $H_0$ = 70 km s$^{-1}$Mpc$^{-1}$, $\Omega_M$ = 0.3, $\Omega_{\Lambda}$ = 0.7. All magnitudes quoted are in the AB system.

\section{Data}\label{data_s}
CLASH \citep{Postman_12} contains 25  galaxy clusters  in the redshift range $0.18< z< 0.89$ which were selected to have temperatures $T_X>$5~keV. Some 20/25 clusters were selected for their X-ray surface brightness symmetry indicating relaxed clusters, and the remaining 5 were specifically selected to have large Einstein radii. The clusters were selected with their lensing properties in mind and at least 18 of the clusters show strong lensing arcs, however the X-ray symmetry selection criteria goes some way to ensure that these clusters are not preferentially aligned along the line-of-sight. We use 23 of these clusters which are described in Table~\ref{data_tab}. The clusters we exclude are MACS1931.8 and MACS0717.5 which both have bright stars near the cluster core, the light from which contaminates the ICL and overall cluster light in a way which we cannot accurately quantify or remove. CLASH contains 16 band HST photometry for each cluster, the data used here were downloaded in reduced form from the CLASH website\footnote{http:\\archive.stsci.edu/prepds/clash/}.

\begin{table*}
\centering
\caption{Clusters from CLASH used in this study.}
\begin{tabular}{l c c c c c}
\hline
Cluster&	 Redshift  & RA	  &Dec	& Cluster mass $M_{200}$  & Cluster mass reference\\	
 		&		&	 &	&	x10$^{15} M_{\odot}$ & (from lensing measurements)		 \\
\hline						
Abell 383 & 0.187 & 02 48 03.36 & -03 31 44.7 & 1.04$\pm$ 0.07 & \citet{Merten_14}\\	
Abell 209 & 0.206 & 01 31 52.57 & -13 36 38.8 & 1.17$\pm$ 0.07 & \citet{Merten_14}\\	
Abell 1423 & 0.213 & 11 57 17.26 & 33 36 37.4 & 1.20$\pm$ 0.59 & \citet{Dahle_06}\\	
Abell 2261 & 0.224 & 17 22 27.25 & 32 07 58.6 & 1.76$\pm$ 0.18 & \citet{Merten_14}\\	
RXJ2129+0005 & 0.234 & 21 29 39.94 & 00 05 18.8 & 0.73$\pm$ 0.18 & \citet{Merten_14}\\	
Abell 611 & 0.288 & 08 00 56.83 & 36 03 24.1 & 1.03$\pm$ 0.07 & \citet{Merten_14} \\	
MS 2137-2353 & 0.313 & 21 40 15.18 & -23 39 40.7 & 1.26$\pm$ 0.06 & \citet{Merten_14}\\	
RXJ1532+30 & 0.345 & 15 32 53.78 & 30 20 58.7 & 0.64$\pm$ 0.09 & \citet{Merten_14}\\	
RXJ2248-4431 & 0.348 & 22 48 44.29 & -44 31 48.4 & 1.40$\pm$ 0.12	& \citet{Merten_14}\\	
MACSJ1115+01 & 0.352 & 11 15 52.05 & 01 29 56.6 & 1.13$\pm$ 0.10	& \citet{Merten_14}\\	
MACSJ1720+35 & 0.391 & 17 20 16.95 & 35 36 23.6 & 0.88$\pm$ 0.08	& \citet{Merten_14}\\	
MACSJ0416-24 & 0.396 & 04 16 09.39 & -24 04 03.9 & 2.50$\pm$ 0.50	& \citet{Zitrin_13}\\	
MACSJ0429-02 & 0.399 & 04 29 36.10 & -02 53 08.0 & 0.96$\pm$ 0.14 & \citet{Merten_14}\\	
MACSJ1206-08 & 0.440 & 12 06 12.28 & -08 48 02.4 & 1.00$\pm$ 0.11	& \citet{Merten_14}\\	
MACSJ0329-02 & 0.450 & 11 57 17.26 & -02 11 47.7 & 0.86$\pm$ 0.11 & \citet{Merten_14}\\	
RXJ1347-1145 & 0.451 & 13 47 30.59 & -11 45 10.1 & 1.35$\pm$ 0.19	& \citet{Merten_14}\\	
MACSJ1311-03 & 0.494 & 13 11 01.67 & -03 10 39.5 & 0.53$\pm$ 0.04 & \citet{Merten_14}\\	
MACSJ1149+22 & 0.544 & 11 49 35.65 & 22 23 55.0 & 5.10$\pm$ 1.90	& \citet{Sereno_12}\\
MACSJ1423+24 & 0.545 & 14 23 47.76 & 24 04 40.5 & 0.65$\pm$ 0.11 & \citet{Merten_14}\\	
MACSJ2129-07 & 0.570 & 21 29 26.06 & -07 41 28.8 & 3.50$\pm$ 3.10	& \citet{Sereno_12}\\
MACSJ0647+70 & 0.584 & 06 47 50.03 & 70 14 49.7 & 6.80$\pm$ 1.40	& \citet{Sereno_12}\\
MACSJ0744+39 & 0.686 & 07 44 52.80 & 39 27 24.4 & 0.79$\pm$ 0.04	& \citet{Merten_14}\\	
CLJ1226+3332 & 0.890 & 12 26 58.37 & 33 32 47.4 & 1.72$\pm$ 0.11	& \citet{Merten_14}\\

\hline
\end{tabular}
\label{data_tab}
\end{table*}

Since the ICL is extended and has low surface brightness it could be easily mistaken for sky background during data reduction and subtracted. \citet{Presotto_14} examine one of the CLASH clusters (MACSJ1206, $z=0.44$) using multi-band SUBARU data, and as part of their data reduction they test that sky subtraction does not cause the ICL to be removed by comparing their Subaru reduction with the HST data used here. They find the ICL to be detected to the same surface brightness levels in both data sets. Since both data use independent sky subtraction methods this shows that the over-subtraction of ICL which is mistaken for sky is not an issue here.

\section{Analysis}\label{analysis_section}
Before we analyse the data we first remove any point sources and foreground or background galaxies from the images. Point sources and non-cluster member galaxies are identified using the object catalogues for the CLASH sample described in \citet{Jouvel_14}. Point sources are defined from the CLASH catalogues as objects having a ``stellarity'' value of 0.1 or greater. In line with \citet{Jouvel_14}, a stellarity less than 0.08 selects 86\% of objects in any given CLASH cluster image -  a value of 0.1 effectively identifies all the galaxies in an image. The point sources in the data were thus identified and were masked out  using a circle of area 1.5 times the area defined for that object in the catalogue. The data were then visually examined and any remaining obvious point sources were removed by hand. Cluster member galaxies are identified as any object whose photometric redshift from the photo-z catalogue \citep[][]{Jouvel_14} is within $3\sigma$ of the cluster redshift, where the $1\sigma$ accuracy of the photometric redshifts is defined in \citet{Jouvel_14} as $0.04\times(1+z_{object})$.  Non-cluster galaxies were also masked out using a circle of 1.5 times the area given in the catalogue, in the case of gravitational lens arcs, the arcs were masked out by hand.

\subsection{SED fitting and stellar mass measurements}\label{bcg_mass}

To measure the stellar masses of the BCGs and galaxies within 50\,kpc of the BCG in the CLASH sample we fit their spectral energy distributions (SEDs) using photometry from the CLASH ACS (for this study we used the F606W, F625W, F775W, F814W, and F850LP bands) and WFC3 (F105W, F110W, F125W, F140W, F160W) images. Magnitudes were measured in each band using the {\tt MAG\_AUTO} method in {\tt SExtractor} (\citealt{BertinArnouts96}; similar to the method described in \citealt{Stott_10}), using consistent aperture sizes between wavebands for each galaxy, set by the F140W band. 

We fit the resulting SEDs using a similar methodology to that described in \citet[][see also \citealt{Hilton_2010, Menanteau_2012}]{Shapley_2005}. We construct a grid of \citet{BruzualCharlot_2003} solar metallicity models with exponentially declining star formation histories with 20 values of $\tau$ in the range $0.1-20$~Gyr, and 172 ages in the range $0.001-13.0$~Gyr. We adopt a \citet{Salpeter_1955} initial mass function (IMF), and model the effect of dust extinction using the \citet{Calzetti_2000} law, allowing $E(B-V)$ to vary in the range $0.0-0.48$ in steps of 0.04. We use $\chi^2$ minimisation to find the best fit and determine the stellar mass of each galaxy from the normalisation of the best fit model template to the observed SED. We restrict the range of model templates to those with ages less than the Universe at the redshift of each cluster. Despite the fact that degeneracies exist between model parameters (e.g., age, $\tau$, $E(B-V)$; see the discussion in \citealt{Shapley_2005}), many authors have shown that stellar masses measured from SED fitting are robust to variations in model parameters \citep[e.g.][]{ForsterShrieber_2004, Shapley_2005}. We note that uncertainties in the IMF and the contribution of thermally pulsating AGB stars to infrared fluxes \citep[e.g.][]{Maraston_2005, Conroy_2009} is likely to lead to the stellar mass estimates only being accurate to within a factor of $\sim 2$.

\subsection{ICL Measurement}\label{icl_method}
To measure the ICL we follow the method described in \citet{B12} which is summarised here. The total cluster light is measured by summing all the flux in the images after non-cluster objects have been masked out in the manner described above. The fraction of the total cluster light contained in the ICL is simply estimated by summing all the light below a given surface brightness threshold and comparing this to the total cluster light. In this paper we use a threshold of 25 mag/arcsec$^2$ in the rest-frame B-band which is comparable to that used in previous studies \citep[e.g.,][]{Feldmeier_04, Krick_06, Krick_07, B12, Presotto_14}. 
Whilst this method is very simplistic it avoids the necessity of disentangling the low brightness light at the edges of galaxies from the faint ICL, which is impossible to do definitively without dynamical information about the stars in the cluster. This method also avoids the necessity of fitting both the ICL and BCG together with a surface brightness profile \citep[e.g., see][]{Zibetti_05, Gonzalez_07, Gonzalez_13, Toledo_11}. The simplicity of this method also makes it very effective for comparing ICL results across different studies and data samples (e.g., above references). See \citet{B12} for a more detailed discussion of this method.

\begin{figure*}
\includegraphics[width=6.5cm]{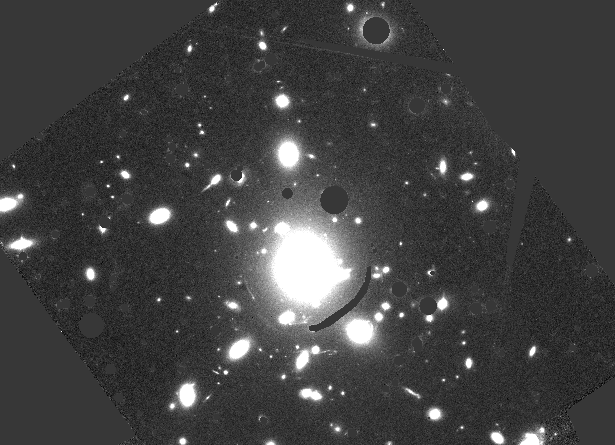}
\includegraphics[width=6.5cm]{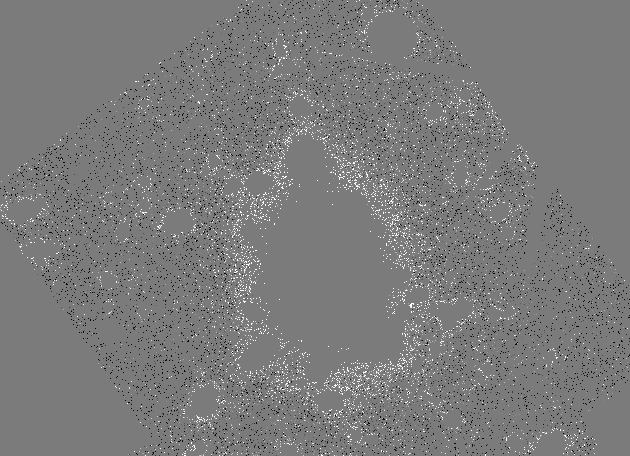}
\includegraphics[width=6.5cm]{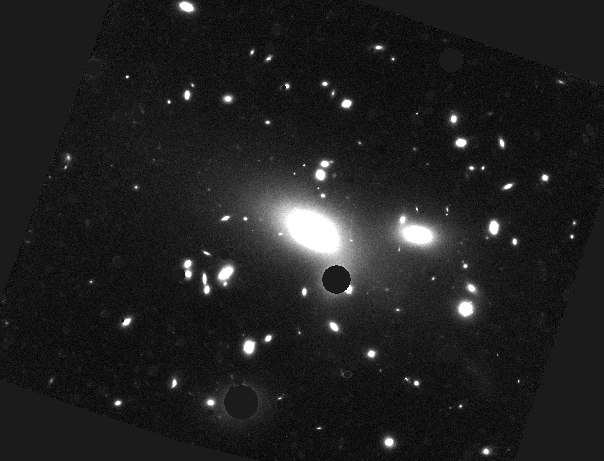}
\includegraphics[width=6.5cm]{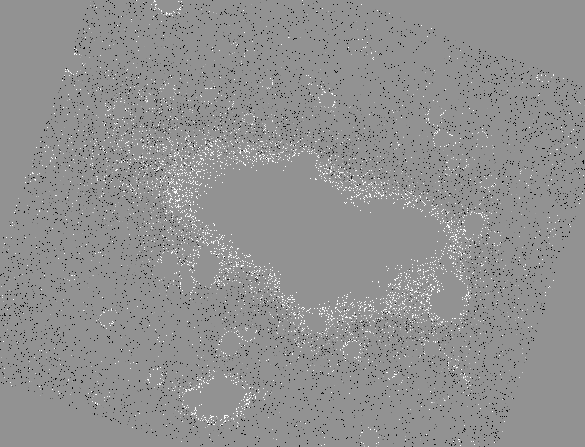}
\caption{Left: The central regions of Abell~383 (top) and Abell~1423 (bottom). Right: the same regions with all pixels except those counted as ICL masked out, equivalent to a surface brightness threshold of 25 mag/arcsec$^2$ in the rest-frame B-band. Up is North, left is East. Images are approximately 2'50'' on each side corresponding to 525~kpc for Abell~383 and 575~kpc for Abell~1423.}
\label{icl_image}
\end{figure*}

Due to the clusters being at a variety of redshifts, the rest frame B-band needs to be converted to the appropriate observed frame waveband from which the ICL is to be measured. This is done using the standard redshift-wavelength equation $\lambda_{obs}=\lambda_{emit}/(1+z)$, the observed wavebands for each cluster are listed in Table~\ref{corr_tab}. The rest-frame 25 mag/arcsec$^2$ surface brightness limit also needs to be converted to an observed surface brightness limit for each individual cluster by applying the redshift surface-brightness dimming correction ($2.5\log(1+z)^3$) and the K-correction for each cluster's respective redshift. The K-correction calculated for these observations is a differential K-correction, $dK(z)$, taking into account the difference in central wavelength of the rest-frame B-band and the de-redshifted observed frame wave-band. The $dK$ values were calculated using the stellar population synthesis models of \citet{BruzualCharlot_2003} assuming an old stellar population with a formation redshift of $z_f=3$ and a solar metallicity. This is obtained using the magnitudes derived from the \citet{BruzualCharlot_2003} models as, 
\begin{equation}
dK(z) = M_{ev}(z)-M_{rf}(z) + 2.5\log (1+z),
\end{equation}
where the $M_{ev}$ term is from the observed HST filter, and the $M_{rf}$ term is from the rest frame B-band. The values for $dK$ are shown in Table~\ref{corr_tab}. The observed band surface brightness limit is thus given by, 
\begin{equation}
\mu_{obs}(z) = \mu_{rest}(z) +2.5\log(1+z)^3 + dK(z),
\end{equation}
(see \citealt{Hogg_02}). The corrected values for the observed surface brightness threshold are given in Table~\ref{corr_tab}. Examples of the cluster cores with the ICL thresholds applied are shown in Figure~\ref{icl_image}.  The $2\sigma$ surface brightness limit for the data is 26.3, 26.4, 25.7 and 25.5 mag/arcsec$^2$ for the F606W, F625W, F775W and F850LP filters respectively. Since the observed surface brightness thresholds for the F775W and F850LP filters are fainter than the sky limit we cannot measure the ICL below 25 mag/arcsec$^2$ for the higher redshift clusters in CLASH. All the clusters observed in the F625W filter above a redshift of $z=0.4$ are also excluded from the ICL analysis for this reason (see Table~\ref{corr_tab}).

The error on our measurement of the ICL fraction of the total cluster light is estimated from the variance of the flux (f) in the ICL as,
\begin{equation}
error(ICL)=\sqrt{\Big(\frac{\sigma_{ICL}}{f_{total}}\Big)^2  + \Big(\frac{f_{ICL}.\sigma_{total}}{(f_{total})^2}\Big)^2}
\end{equation}
where $\sigma$ denotes the standard deviation.

\begin{table*}
\centering
\caption{The corrections applied for ICL measurement and the observed surface brightness threshold equivalent to the rest frame limit of $\mu_B=$25 mag/arcsec$^2$.}
\begin{tabular}{l c c c c c c}
\hline
Cluster  & z & Observed waveband &  $M_{ev}(z)-M_{rf}(z)$ &  dK  & Surface brightness & Corrected surface brightness \\
	&	&   for ICL measurement &  &	& dimming correction	& equivalent to 25 mag/arcsec$^2$\\
	&	&						&   &	 &		$2.5\log(1+z)^3$		& at $z=z$(cluster)\\
\hline
Abell~383 & 0.187 & F606W & -0.68 & -0.49 & 0.56 & 25.06\\
Abell~209 & 0.206 & F606W & -0.62 & -0.41 & 0.61 & 25.20\\
Abell~1423 & 0.213 & F606W & -0.55 & -0.34 & 0.63 & 25.29\\
Abell~2261 & 0.224 & F606W & -0.55 & -0.33 & 0.66 & 25.33\\
RXJ2129+0005 & 0.234 & F606W & -0.44 & -0.21 & 0.68 & 25.48\\
Abell~611 & 0.288 & F606W & -0.39 & -0.11 & 0.82 & 25.71\\
MS~2137-2353 & 0.313 & F606W & -0.27 & 0.03 & 0.89 & 25.91\\
RXJ1532+30 & 0.345 & F606W & -0.20 & 0.12 & 0.97 & 26.09\\
RXJ2248-4431 & 0.348 & F606W & -0.20 & 0.13 & 0.97 & 26.10\\
MACSJ1115+01 & 0.352 & F606W & -0.20 & 0.13 & 0.98 & 26.11\\
MACSJ1720+35 & 0.391 & F625W & -0.23 & 0.13 & 1.07 & 26.21\\
MACSJ0416-24 & 0.396 & F625W & -0.23 & 0.14 & 1.09 & 26.22\\
MACSJ0429-02 & 0.399 & F625W & -0.23 & 0.14 & 1.09 & 26.23\\
MACSJ1206-08 & 0.440 & F625W & -0.01 & 0.39 & 1.19 & 26.58\\
MACSJ0329-02 & 0.450 & F625W & -0.01 & 0.40 & 1.21 & 26.61\\
RXJ1347-1145 & 0.451 & F625W & -0.01 & 0.40 & 1.21 & 26.61\\
MACSJ1311-03 & 0.494 & F625W & 0.25 & 0.69 & 1.31 & 27.00\\
MACSJ1149+22 & 0.544 & F625W & 0.49 & 0.96 & 1.41 & 27.37\\
MACSJ1423+24 & 0.545 & F775W & -0.78 & -0.30 & 1.42 & 26.11\\
MACSJ2129-07 & 0.570 & F775W & -0.73 & -0.24 & 1.47 & 26.23\\
MACSJ0647+70 & 0.584 & F775W & -0.68 & -0.18 & 1.50 & 26.31\\
MACSJ0744+39 & 0.686 & F775W & -0.50 & 0.07 & 1.70 & 26.77\\
CLJ1226+3332 & 0.890 & F850LP & -0.89 & -0.20 & 2.07 & 26.88\\
\hline
\end{tabular}
\label{corr_tab}
\end{table*}

\subsection{Mergers onto the BCG}\label{nc_method}
In order to estimate the stellar mass growth of the BCG and ICL we estimate the number of mergers which will occur with the BCG. To do this we follow the method described in \citet{BC13} (hereafter BC13) of counting the number of companions to the BCG ($N_c$) within a fixed aperture. For comparison with previous studies we use an aperture of 50~kpc, centred on the BCG. The BCG companions were detected using SExtractor \citep{BertinArnouts96}.

We carried out completeness tests for the detection of BCG companions by inserting fake model galaxies, with \citet{DeVauc} profiles and half-light radii of 5~kpc (a similar size to the smaller BCG companions in the data), into the cluster images at random positions within a 100~kpc radius of the BCG. We then ran SExtractor over the images and counted the number of model galaxies that were detected. We decreased the brightness of the model galaxies inserted until the detection completeness fell below 90\%. We found our detection of companions to be complete to a minimum brightnesses of 1:1000 of the BCG. All companions  subsequently counted in our results have brightnesses above this limit; any companions fainter than this limit are excluded from our analysis.
The companions counted by SExtractor were also checked by eye to ensure that bright companions near to the BCG were not accidentally excluded and close pairs of objects were not miscounted.

In order to be able to say whether or not a BCG companion will merge with the BCG in the time between the cluster redshift and the present we need to know its merging timescale. As described in BC13, we estimate the amount of time taken for a companion on a circular orbit to be accreted onto the BCG using the dynamical friction timescale, $T_{fric}$ (Gyrs), which is given by, 

\begin{equation}
T_{fric}= 3.81\times10^5 ~ \frac{r^2v_c}{\mathcal{M}_{c}}.
\end{equation}
Here $r$ is the separation of the galaxies in kpc, $v_c$ is the circular velocity of the galaxy cluster in km\,s$^{-1}$, and $\mathcal{M}_c$ is the total mass of the companion in solar masses (for a more thorough description see \citealt{BinneyTremaine}; and for a more detailed description of its use in this instance see BC13).

We estimate the stellar masses of the BCGs and their companions by fitting their SED in the same manner as described above. In order to calculate their total masses ($\mathcal{M}_{c}$) we then assume a mass-to-light ratio of M/L=4 (in the rest-frame B-band), in line with BC13 and references therein. In the absence of dynamical information, the circular velocity of the companions is also estimated in the same way as BC13. Typical values of circular velocities in galaxy cluster cores range between $v_c\sim350$~km\,s$^{-1}$ and $v_c\sim700$~km\,s$^{-1}$ (see BC13). In order to calculate $T_f$ we use an average of these and estimate the circular velocity to be $v_c=550$~km\,s$^{-1}$. These values for M/L and $v_c$ combined with the stellar masses measured from the SED fitting give the typical dynamical friction timescales, $T_{fric}$, and are summarised in Table~\ref{tf_tab}. As is clear in Table~\ref{tf_tab} the least massive BCG companions have merging timescales in excess of the Hubble time. The dynamical fraction timescale was calculated for each BCG companion individually, whose magnitude is above the completeness limit. Only companions which have dynamical friction timescales less than the time between the observed cluster redshift and the present day are counted as contributing to the BCG mass growth in the subsequent analysis.

\begin{table}
\centering
\caption{Dynamical friction timescales for typical masses of companions measured from SED fitting assuming a mass-to-light ratio of M/L=4, $v_c$=550 km\,s$^{-1}$ and an initial distance from the BCG of $r$=50~kpc.}
\begin{tabular}{c c  }
\hline
Stellar mass ratio		& $T_{fric}$ (Gyr) \\
\hline
1:2			&0.38\\
1:5			&0.96\\
1:10			&1.93\\
1:20			&3.86\\
1:50			&9.65\\
1:100		&19.29\\
1:1000		&192.93\\
\hline
\end{tabular}
\label{tf_tab}
\end{table}

The errors on the numbers of BCG companions are calculated using the photometric redshift accuracy quoted in \citet{Jouvel_14}. Jouvel et al. find 96\% of the CLASH cluster galaxies with spectroscopic redshifts lie within 2$\sigma$ of their photometric redshift. Thus we would expect the percentage of companions which would be misidentified non-cluster members (or cluster members which would be erroneously excluded) to be 4\% of the number of companions counted, thus the error on $N_c$ is 4\% of $N_c$.

\section{Results}\label{results_s}
The results for BCG masses, ICL fractions, numbers of BCG companions ($N_c$) and inferred mass growth from their accretion are summarised in Table~\ref{results_tab}.

\subsection{Results for BCG masses}

\begin{figure}
\centering
\includegraphics[width=8.5cm]{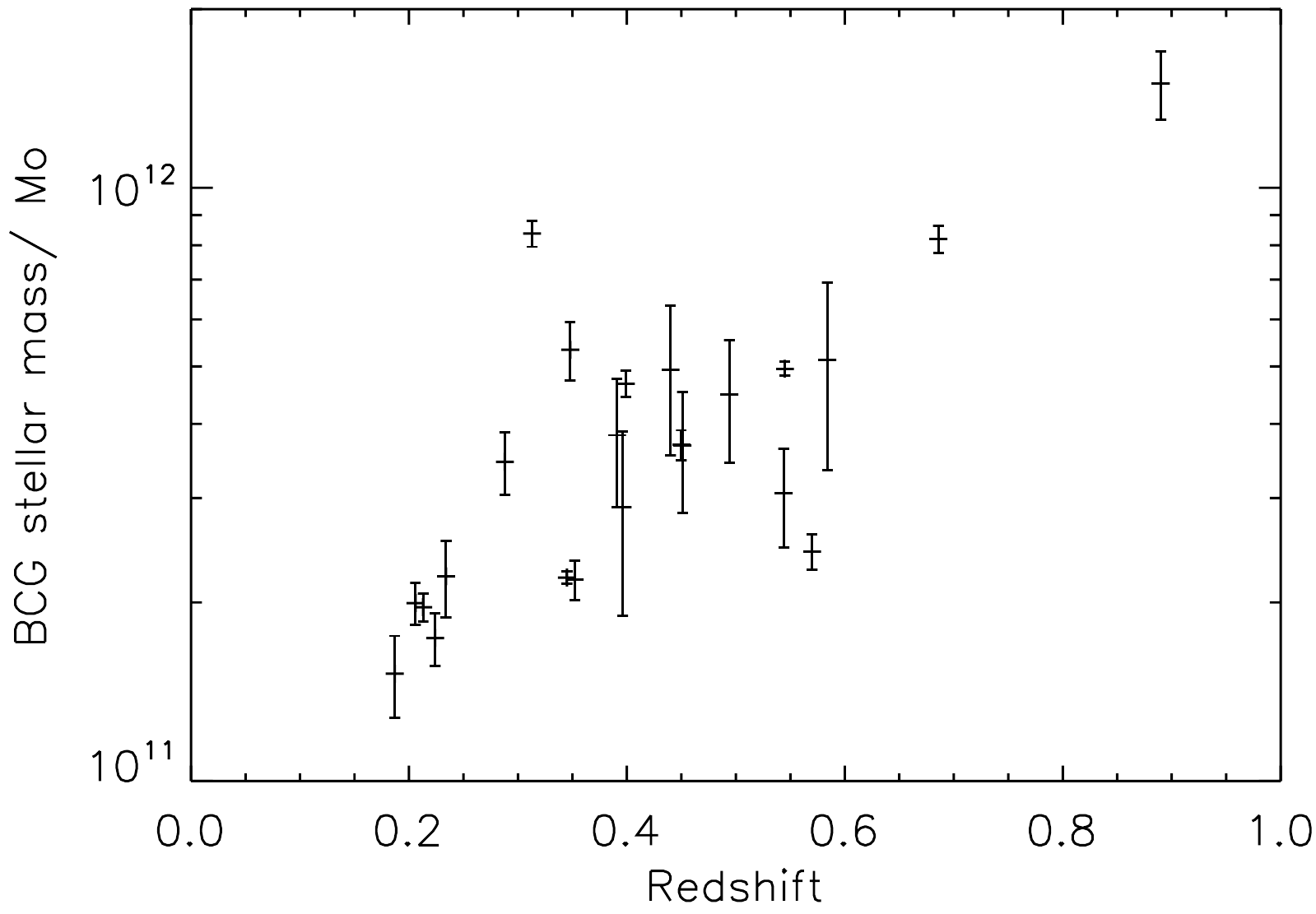}
\caption{Relation between BCG stellar mass and redshift.}
\label{BCG_z}
\end{figure}

\begin{figure}
\centering
\includegraphics[width=8.5cm]{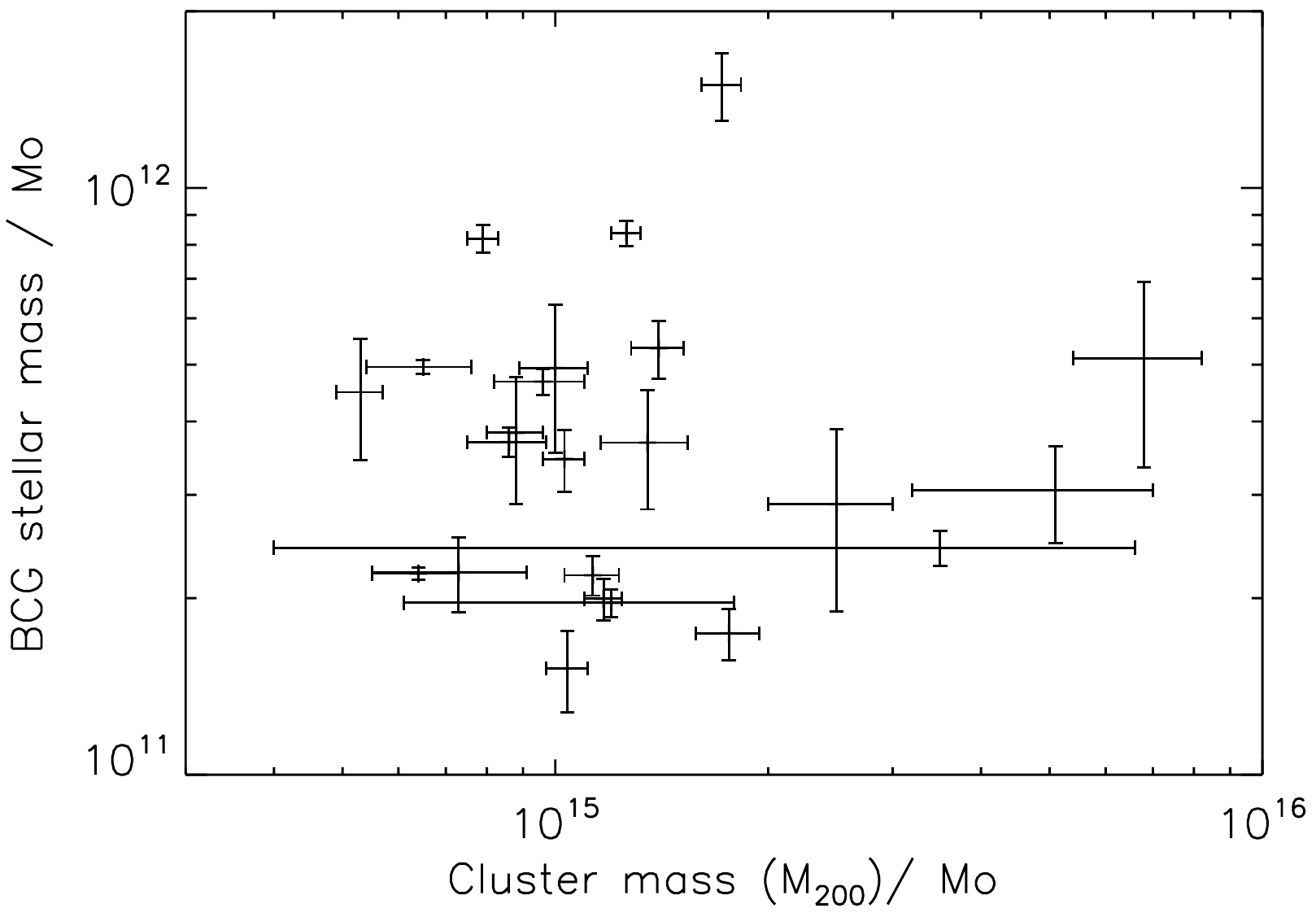}
\caption{BCG stellar mass vs cluster total mass ($M_{200}$), see Table~\ref{data_tab} for origin of cluster masses.}
\label{bcg_cl_mass}
\end{figure}

\begin{figure}
\centering
\includegraphics[width=8.5cm]{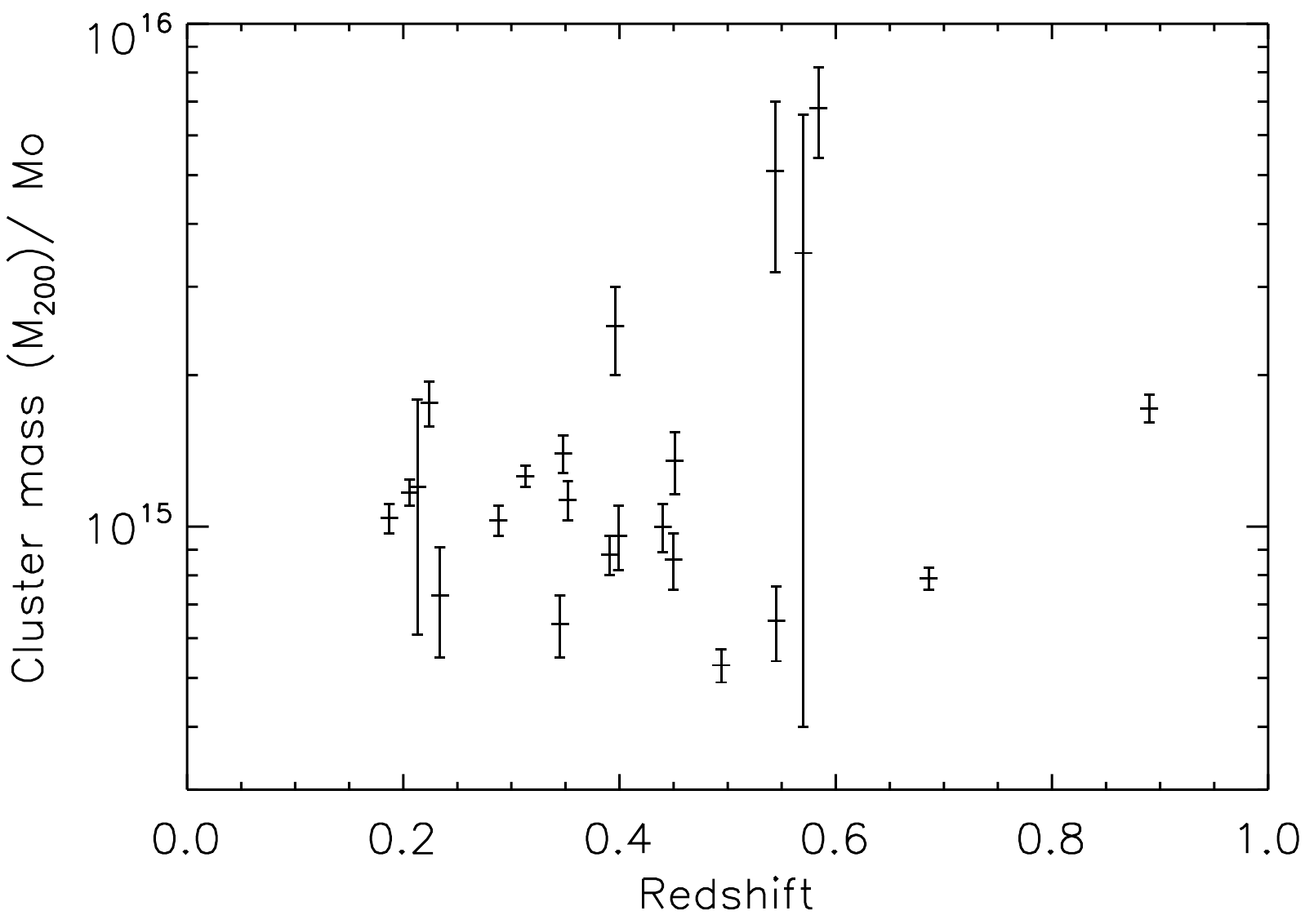}
\caption{Cluster mass ($M_{200}$) vs redshift for CLASH.}
\label{cl_mass}
\end{figure}

 We will define the level of correlation for our results using the Spearman rank correlation. The Spearman rank coefficient (SR) is between zero and one with SR=0 for highly uncorrelated variables and SR=1 (--1) for highly correlated (anti-correlated) variables. The Spearman rank also includes a significance of deviation from zero (SDZ) indicating the statistical strength of the correlation. SDZ represents the probability that a value greater than or equal to the observed SR would be calculated under the null hypothesis (of no correlation), and is given by SDZ=SR($\sqrt{(n-2)/(1-SR^2)}$), where $n$ is the number of degrees of freedom, and SDZ is distributed as a Student's t-distribution with $n-2$ degrees of freedom. SDZ has values between 0 and 1 where a small value for SDZ indicates a high significance for SR. 
 
Figure~\ref{BCG_z} shows the relation between the BCG stellar masses and redshift, the Spearman rank coefficient for this figure is SR= 0.7 and its statistical significance is SDZ= 10$^{-3}$, indicating a moderately strong correlation with a high statistical significance. This figure clearly shows that the two highest redshift clusters house the most massive BCGs in the CLASH sample. Even without these two high redshift, high mass BCGs, the rest of the sample still forms a fairly significant correlation (SR=0.6, SDZ=10$^{-2}$). However when the two highest redshift points are excluded the slope of the gradient for a simple $y=mx+c$ fit decreases by half, forming a fairly flat relation. 
Figure~\ref{bcg_cl_mass} clearly shows that there is no correlation between cluster mass and BCG stellar mass, with a Spearman rank coefficient SR= $-0.05$ and SDZ= 0.8. Given that cluster masses and BCG masses are usually seen to be positively correlated \citep[e.g.,][]{Stott_10}, this suggests that the almost flat trend seen here is due to a selection bias of the CLASH sample, whereby clusters with BCGs of fairly narrow mass have been selected. The relation between cluster mass and redshift is shown in Figure~\ref{cl_mass}, where no correlation is seen (SR=0.1 and SDZ=0.6). The majority of the clusters in the sample (18/23) fall over a narrow range in cluster mass ($6\times10^{14}M_{\odot}-2\times10^{15}M_{\odot}$) so perhaps it is not surprising to see the BCG masses showing little variation with redshift (with 18/23 BCGs being in a $5\times10^{11}M_{\odot}$ mass range). We suggest that the BCG masses recorded for CLASH are a result of selection, rather than providing evidence for lack of BCG mass growth or an inverted BCG mass growth. The absence of any correlation between cluster mass and redshift means subsequent trends should not be affected by the variation of cluster mass with redshift.

\subsection{Results for ICL fractions}

\begin{figure}
\centering
\includegraphics[width=8.5cm]{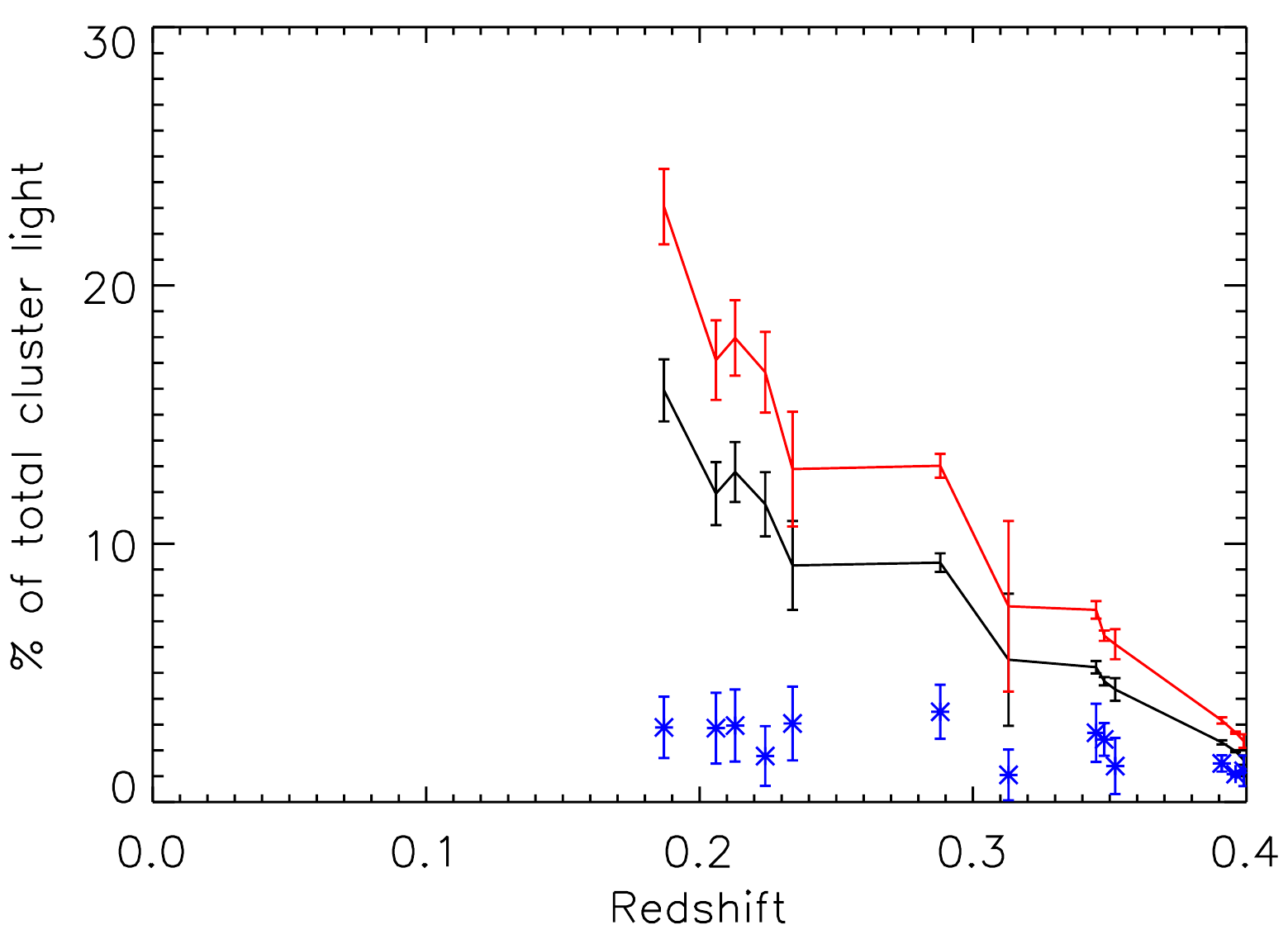}
\caption{ Fraction of total cluster light contained in the ICL below a surface brightness threshold of observed frame 25 mag/arcsec$^2$ and above lower limits of 25.5 mag/arcsec$^2$ (black) and 26 mag/arcsec$^2$ (red); and the fraction of the total cluster light contained in the BCG (blue stars).}
\label{bcg_icl_frac}
\label{ICL_z_vlt}
\end{figure}

Figure~\ref{ICL_z_vlt} shows the fraction of the total cluster light contained in the ICL for the CLASH sample. As mentioned above the higher redshift ($z>0.4$) clusters in the sample have cosmology corrected, observed 25 mag/arcsec$^2$ surface brightness thresholds which are fainter than the $2\sigma$ sky noise level and are thus excluded from any ICL analysis - we limit our ICL analysis to below this redshift.

A trend of increasing ICL fraction with decreasing redshift is clearly seen, with a strong correlation of SR=$-0.99$ and a high significance of SDZ= 10$^{-10}$ (for both black and red points). The CLASH sample shows an increase in the amount of the total cluster light contained in the ICL of a factor of $\sim4-5$ between $z=0.40$ and $z=0.19$.

The $2\sigma$ sky background levels for the CLASH data are $\sim25.5-26.4$mag/arcsec$^2$, (see Section~\ref{icl_method}) meaning the ICL is only measured above the sky for clusters at redshift $z<0.4$. As redshift increases, cosmological dimming and waveband shift effects cause the observed surface brightness equivalent to the rest frame 25 mag/arcsec$^2$ to become fainter. The upper limit on surface brightness measured eventually approaches the level of the sky, making the ICL unmeasurable below 25 mag/arcsec$^2$ at higher redshifts. Even with the limit of $z<0.4$ some of the higher redshift clusters left over have their 25 mag/arcsec$^2$ threshold limit within 0.5 mag/arcsec$^2$ of the sky. In order to test the possible bias introduced by measuring the ICL close to the sky noise limit we measured the ICL over surface brightness ranges of 0.5 and 1.0 mag/arcsec$^2$ below the 25 mag/arcsec$^2$ threshold (25.5 and 26 mag/arcsec$^2$). The results of this are shown in Figure~\ref{ICL_z_vlt}, where the black points indicate the range 25-25.5 mag/arcsec$^2$ and the red points indicate the 25-26 mag/arcsec$^2$ range. We find that a fainter lower limit of 26 compared to 25.5 mag/ arcsec$^2$ increases the ICL fraction measured by an average of 41$\pm$3\% (i.e. increasing the amount of the total cluster light in the ICL from $\sim$16\% to 23\%). Lowering the faint end limit above which the ICL is measured has a strong effect on the amount of ICL recorded. We will discuss the effect of the lower surface brightness limit on the ICL recovered in Section~\ref{discussion}. The steepness of slope of the ICL fraction with redshift recorded for both surface brightness lower limits shows that the ICL really is evolving rapidly from $z=0.4-0.2$.

The higher redshift study of \citealt{B12} (hereafter B12) examined clusters at  0.8$<z<$1.2 using near infrared data and showed a quadrupling in the cluster's ICL fraction between their average redshift $z\sim$1 and the present. 
The slope of the points in Figure~\ref{ICL_z_vlt} shows a fairly steep increase in the ICL fraction between $z=0.40$ and $z=0.19$, indicating a rapid but steady buildup of the ICL between intermediate redshifts and the present. This also supports our previous conclusions in BC13 that if many mergers occur in the cores of clusters, rather than increasing the mass of the BCG, the stellar mass in the merging galaxies ends up contributing to the ICL, allowing it to grow rapidly at late times.

Other findings in the literature for the growth of the ICL over similar redshifts show wide ranging values for ICL fractions. \citet{Zibetti_05} find SDSS clusters at $z=0.2-0.3$ to have 11\% ICL; \citet{Giallongo_14} find 23\% ICL in a cluster at $z=0.4$; \citet{Feldmeier_04} find 8--28\% ICL below a surface brightness threshold of V=26 mag/arcsec$^2$ (similar to that used here) in clusters at $z=0.16-0.19$ and \citet{Krick_07} find 6--22\% ICL below the same threshold for clusters at $z=0.05-0.3$ with an increase in the ICL fraction from 11\% to 14\% over the redshift range studied. Simulations of ICL growth also show large ICL fractions at $z=0$, with typical values in the range of 15--45\%  of the total cluster mass or light in the ICL \citep{Contini_14, Purcell_07, Henriques_10, Murante_07, Puchwein_10, Rudick_11, Budzynski_14}, values which are in general good agreement with the low redshift end of the CLASH data.  Both the observational and simulated results from the literature show no tight consensus on the fraction of total cluster light contained in the ICL at low redshifts ($0<z<0.3$), with measurements in the general range of 10--30\% over a range of cluster masses, wavebands and surface brightness thresholds. The observed ICL fractions found here fit reasonably well (qualitatively) within the range found in other observational studies, even though the clusters and methods of measurement vary between studies. However, the lower surface brightness limits of the measurements in these other studies are not well described, and as discussed above this may significantly affect the amount of ICL measured.

\begin{figure}
\centering
\includegraphics[width=8.5cm]{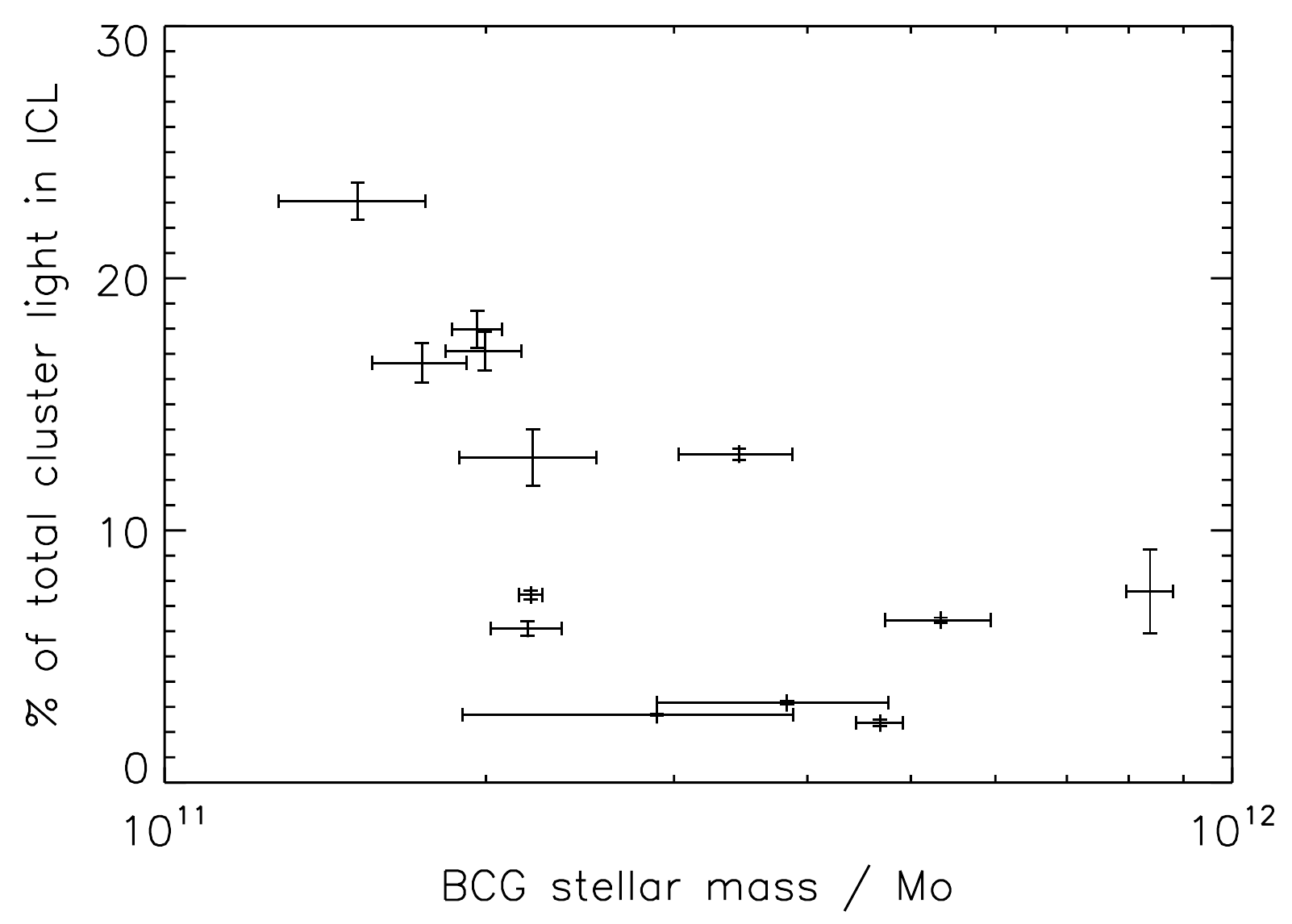}
\caption{The fraction of the total cluster light contained in the ICL for CLASH vs BCG stellar mass.}
\label{bcg_icl}
\end{figure}

In Figure~\ref{bcg_icl} we compare the BCG masses to the ICL fractions. This figure shows an increase in ICL fraction with decreasing BCG mass, with a  fairly strong correlation of SR=$-0.7$, with a high statistical significance SDZ= 0.01. However, the clusters with the highest ICL fractions are the lowest redshift clusters, and the lowest mass BCGs are also at lower redshifts (see Figure~\ref{BCG_z}), suggesting that the correlation seen is purely due to redshift. A partial Spearman rank correlation allows one to determine the correlation between two variables A and B, given that they are both also correlated with a third variable C, allowing one to quantify the correlation between A and B once their independent correlations with C have been removed. The partial Spearman rank coefficient has the form,
\begin{equation}
SR_{AB,C} = \frac{SR_{AB}-SR_{AC}SR_{BC}} {[(1-SR_{AC}^2)(1-SR_{BC}^2)]^{1/2}} .
\end{equation}
As with the standard Spearman rank coefficient, a value close to $\pm1$ indicates very correlated (or anti-correlated) variables and a value close to zero indicates uncorrelated variables. For a more detailed discussion of the partial Spearman rank see \citet{Collins_Mann}. Given that the BCG mass and ICL fraction are both correlated with redshift we carried out a partial Spearman rank analysis for these three variables. The results of this analysis are shown in Table~\ref{sr_tab} alongside their original Spearman rank coefficient. From this table it is clear that the ICL fraction is correlated with redshift, independent of the BCG mass, but the ICL fraction and BCG mass are not at all correlated with each other when the effect of redshift is included. Therefore the significant correlation found for the ICL fraction and BCG mass from the original Spearman rank analysis is actually spurious. The partial Spearman rank for BCG mass vs redshift is now only 0.2 when the strong ICL vs redshift correlation is accounted for.

\begin{table}
\centering
\caption{Results of partial Spearman rank test for the correlation of redshift, BCG stellar mass and ICL fraction.}
\begin{tabular}{c c}
\hline
Spearman Rank & Partial Spearman rank\\
\hline
$R_{BCG~ ICL}=-$0.7	& $R_{BCG ~ICL, z}=$ 0.2\\
$R_{ICL ~z}=-$0.9		&$R_{ICL~ z, BCG}$= $-$0.8\\
$R_{BCG~ z}=$ 0.7	 	&$R_{BCG~ z, ICL}$= 0.2\\
\hline
\end{tabular}
\label{sr_tab}
\end{table}

\begin{figure}
\centering
\includegraphics[width=8.5cm]{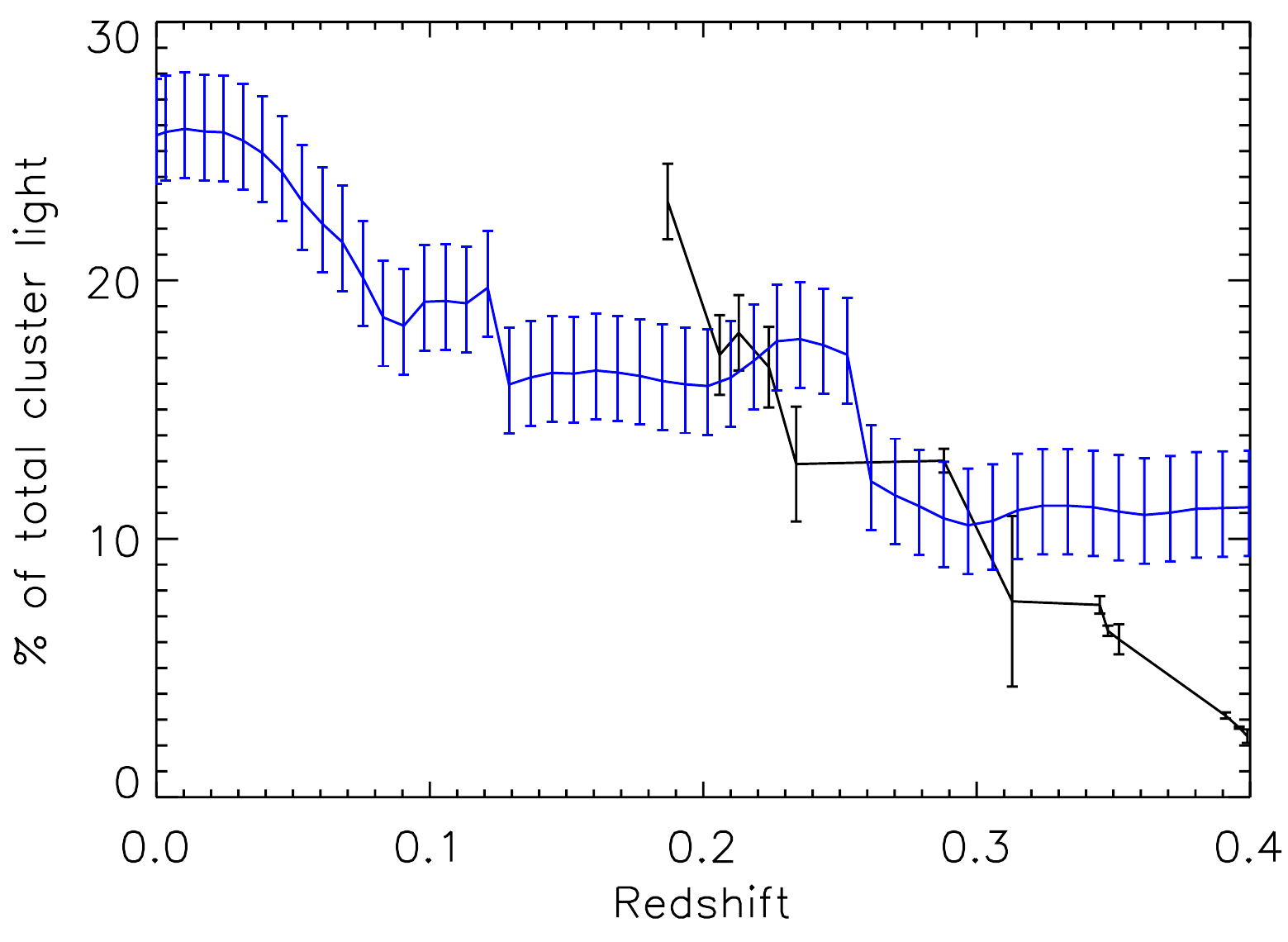}
\caption{Comparison of CLASH (black) ICL fraction measurements with the predictions of the simulations of \citet[blue]{Rudick_11}. The dashed line shows the best fit for the simulated data and the dotted line shows the best fit for the CLASH data.}
\label{rudick_plot}
\end{figure}

Figure~\ref{rudick_plot} shows the ICL fraction with redshift when the CLASH results are compared to the predicted ICL fractions from the simulations of \citet{Rudick_11}. The \citet{Rudick_11} data points are for ICL measured below a surface brightness threshold of 25 mag/arcsec$^2$ in the V-band, which is slightly redder than the observed B-band for CLASH.  In this case one might expect the simulated clusters to have higher ICL fractions as the ICL is thought to be built up from mergers between the red, central galaxies in the core, and in past studies the ICL has been shown to have an old stellar population (e.g., \citealt{Krick_07}).  

The slope of the observed and simulated data clearly differ from each other, with the observed ICL fractions showing much steeper increase with redshift than is predicted by the simulations. One possible cause for this difference may be the depth of the observational data compared to the almost unlimited depth of the simulated data. As discussed above and illustrated in Figure~\ref{bcg_icl_frac}, a change in the lower limit of the flux available to be measured can cause a significant change in the ICL fraction recovered. Since simulated data can probe surface brightness depths at faintness levels well beyond that of observational data it is possible that simulations will predict significantly more ICL than is observed. In the case of \citet{Rudick_11} they are able to measure the ICL down to surface brightnesses of $\mu_V=$35 mag/arcsec$^2$, this level is far below the sensitivity of most current observational data, and the amount and extent of extra flux which may be detected at these faint surface brightness levels is currently unknown.
 
The simulations of \citet{Rudick_11} have cluster masses which are generally smaller than those of the CLASH sample, approximately by an order of magnitude in most cases. However for the order of magnitude range of cluster masses studied by \citet{Rudick_11} there is very little variation in the ICL fraction (see their Figure~4) making the comparison between these simulations and CLASH valid.
 Other simulations of the assembly of the ICL generally show a substantial growth over the redshift range $z=1-0$, \citet{Contini_14} predict a factor of 6 growth in the ICL and \citet{Murante_07} predict factor of 3--5 times growth for the ICL, both of which are similar to the growth found here (factor of 4--5) and in \citet[factor of 4 from $z=1.2$ to $z=0$]{B12}.

\begin{figure}
\centering
\includegraphics[width=8.5cm]{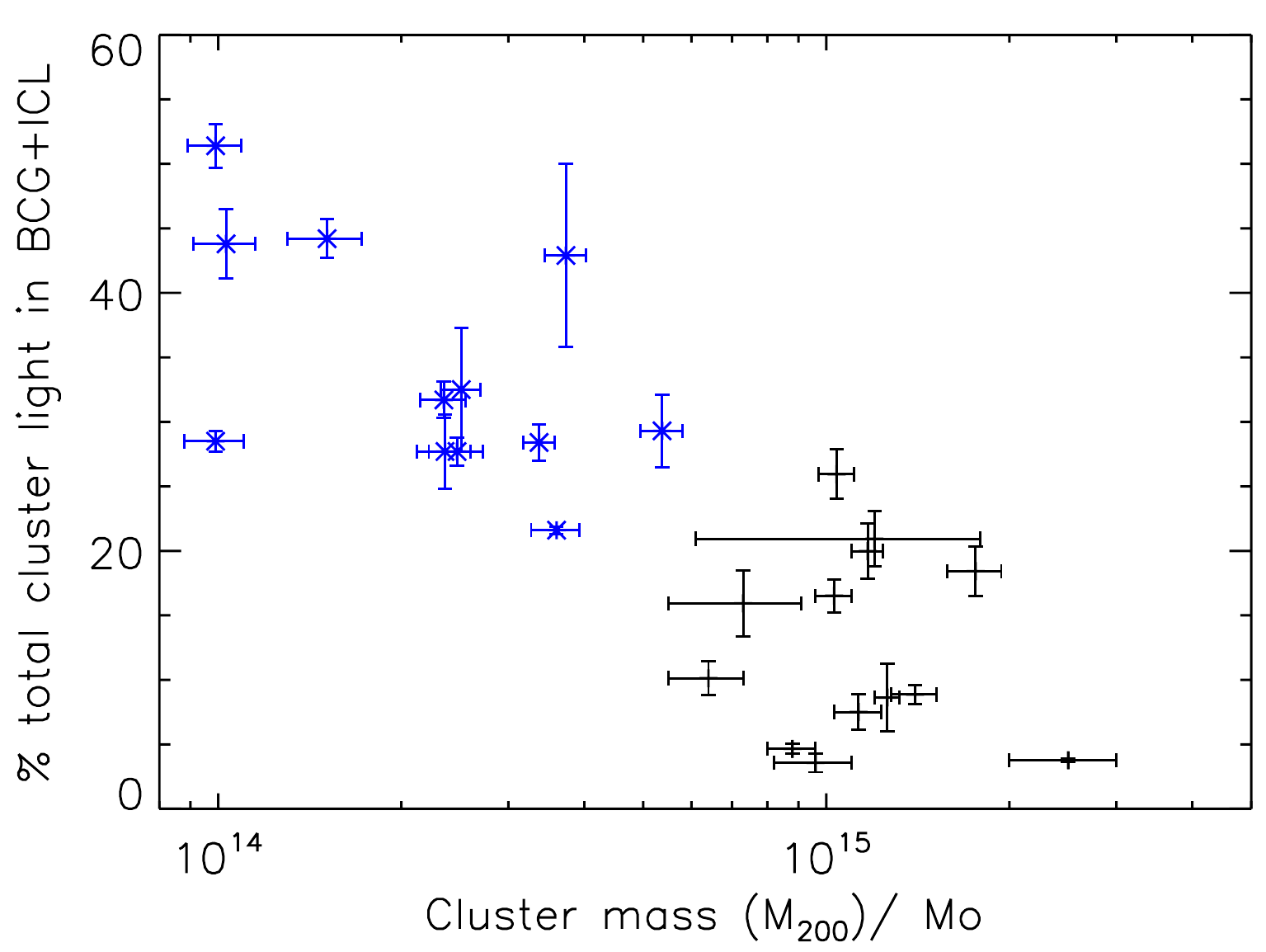}
\caption{Fraction of total cluster light contained in the BCG+ICL against cluster total mass. Black points are from CLASH, blue crosses are the results of \citet{Gonzalez_13} at an average redshift of $z=0.1$.}
\label{bcg+icl}
\end{figure}

When we compare the fraction of the total cluster light that is contained in the ICL with the mass of the host clusters we find a fairly flat distribution, with a large scatter in ICL fraction especially at the lower mass end. No strong trend is found between ICL fraction and cluster mass (SR= -0.1 and SDZ= 0.8). \citet{Krick_06} and \citet{Krick_07} also find no trend between cluster mass and ICL fraction in their $z=0.05-0.3$ Abell clusters. This lack of correlation is also generally found in simulations, where there is no dependence between the fraction of total cluster light or mass contained in the ICL and the host halo mass \citep{Contini_14, Henriques_10, Puchwein_10, Rudick_11}, however in the simulations by \citet{Purcell_07} and \citet{Murante_07} a positive correlation between halo mass and ICL fraction is seen.

The fraction of the total cluster light which is contained in the BCG was measured from the same aperture magnitudes which were used to determine the stellar mass. The fraction of the total cluster light contained within the BCG and ICL for the CLASH sample is shown as the blue points in Figure~\ref{bcg_icl_frac}. From this figure it is clear that the majority of the BCG+ICL light is in the ICL for the CLASH clusters at redshifts $z<0.4$, with the ICL containing 70--80\% of the BCG+ ICL light at these redshifts. This result is in line with the findings of previous studies which fit the surface brightness profile of the BCG+ICL \citep[e.g.,][]{Gonzalez_07}.  At redshift $z\sim0.4$ the ICL and BCG both contain similar fractions of the total cluster light. After $z\lesssim0.4$ the ICL grows rapidly but the BCG stays fairly constant in terms of its fraction of the total cluster light. This indicates that the BCG has assembled the majority of its stellar mass by $z\sim0.4$ at the latest, and the major evolution or merging events after this point are causing the growth of the ICL rather than the BCG.

The BCG+ICL light as a percentage of the total cluster light is plotted against the total cluster mass in Figure~\ref{bcg+icl}, with the results of \citet{Gonzalez_13} for comparison. In this figure there is no trend with mass for the CLASH results on their own (SR= 0.1, SDZ=0.8). Previous studies find that lower mass clusters have higher BCG+ICL fractions \citep[e.g.,][]{Gonzalez_07, Gonzalez_13, Toledo_11}. Over the cluster mass range $0.9\times10^{14}-6\times10^{14} M_{\odot}$ \citet{Gonzalez_13} report BCG+ICL fractions of 20--50\%, similar to the BCG+ICL fractions found for the low mass end of the present study, however the lower mass points of \citet{Gonzalez_13} have a lower average redshift than the CLASH clusters of $z=0.1$ (c.f. $z=0.3$ for CLASH) and therefore may have higher ICL fractions due to having had more time for the clusters to grow their ICL. When the CLASH and Gonzalez points are considered together they form a strong correlation with SR=--0.8, and high significance, SDZ= 10$^{-6}$. This correlation could be due to the changing fraction of ICL between clusters of different masses or it could be due to the evolution of the ICL with redshift.

\subsection{Results for BCG companions and mergers}

\begin{figure}
\centering
\includegraphics[width=8.5cm]{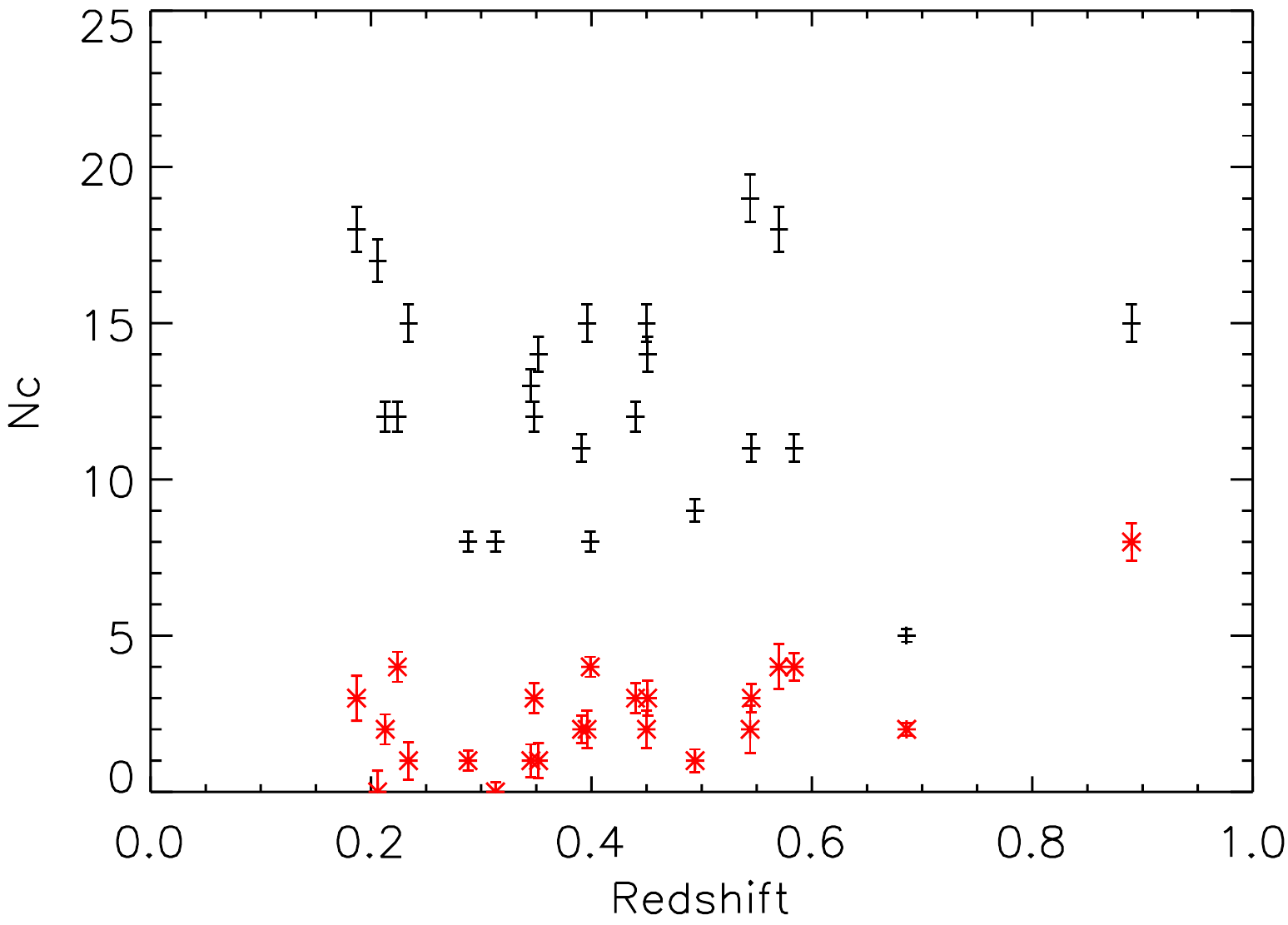}
\includegraphics[width=8.5cm]{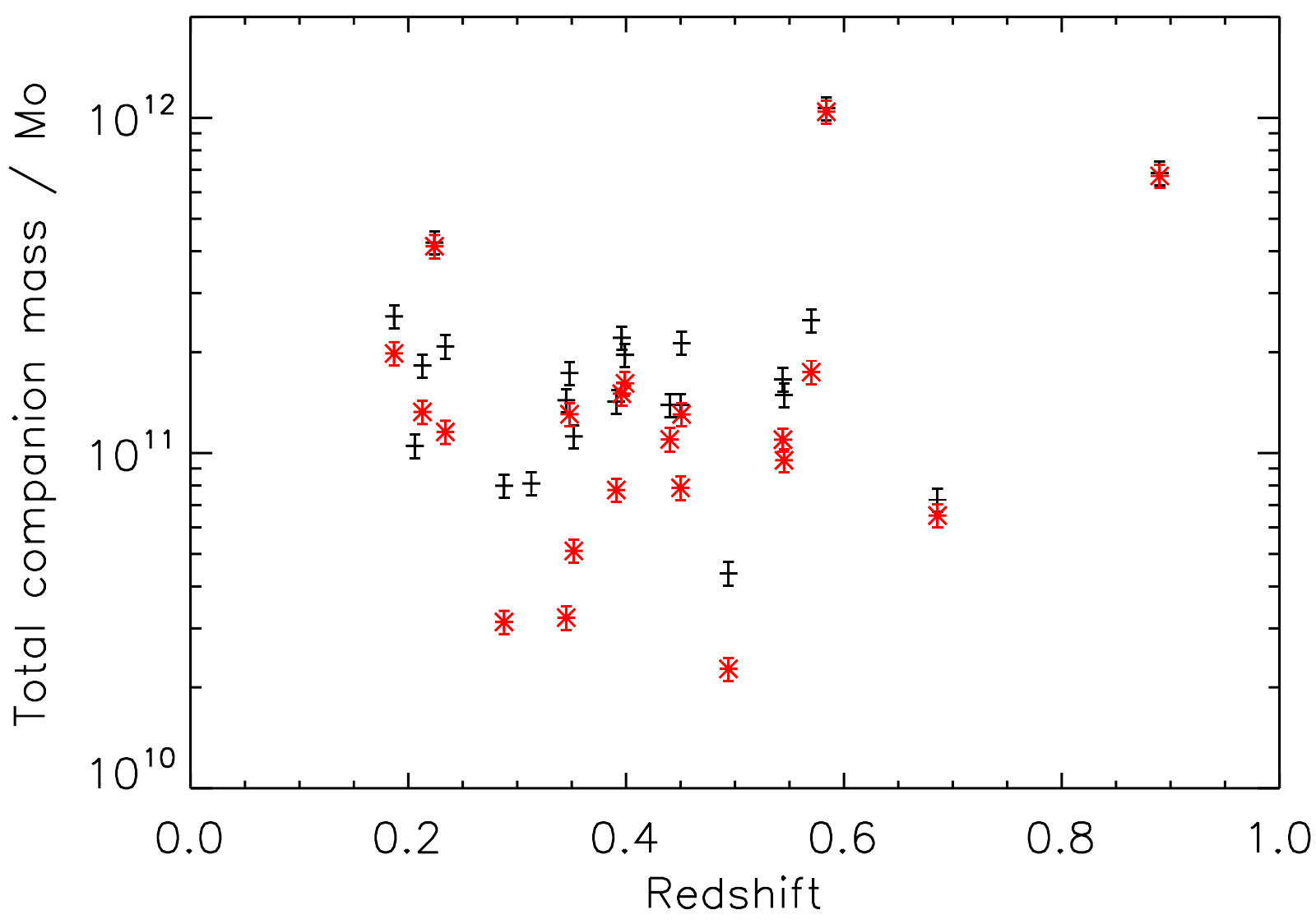}
\caption{Top: The numbers of BCG companions ($N_c$) within a 50~kpc aperture centred on the BCG against redshift. Bottom: Total mass of BCG companions within 50 kpc as a function of redshift. For both plots black points show raw total companion numbers or masses, red crosses show just those companions which will merge with the BCG according to their dynamical friction timescale.}
\label{comp_z}
\label{nc_z}
\end{figure}

\begin{figure}
\centering
\includegraphics[width=8.5cm]{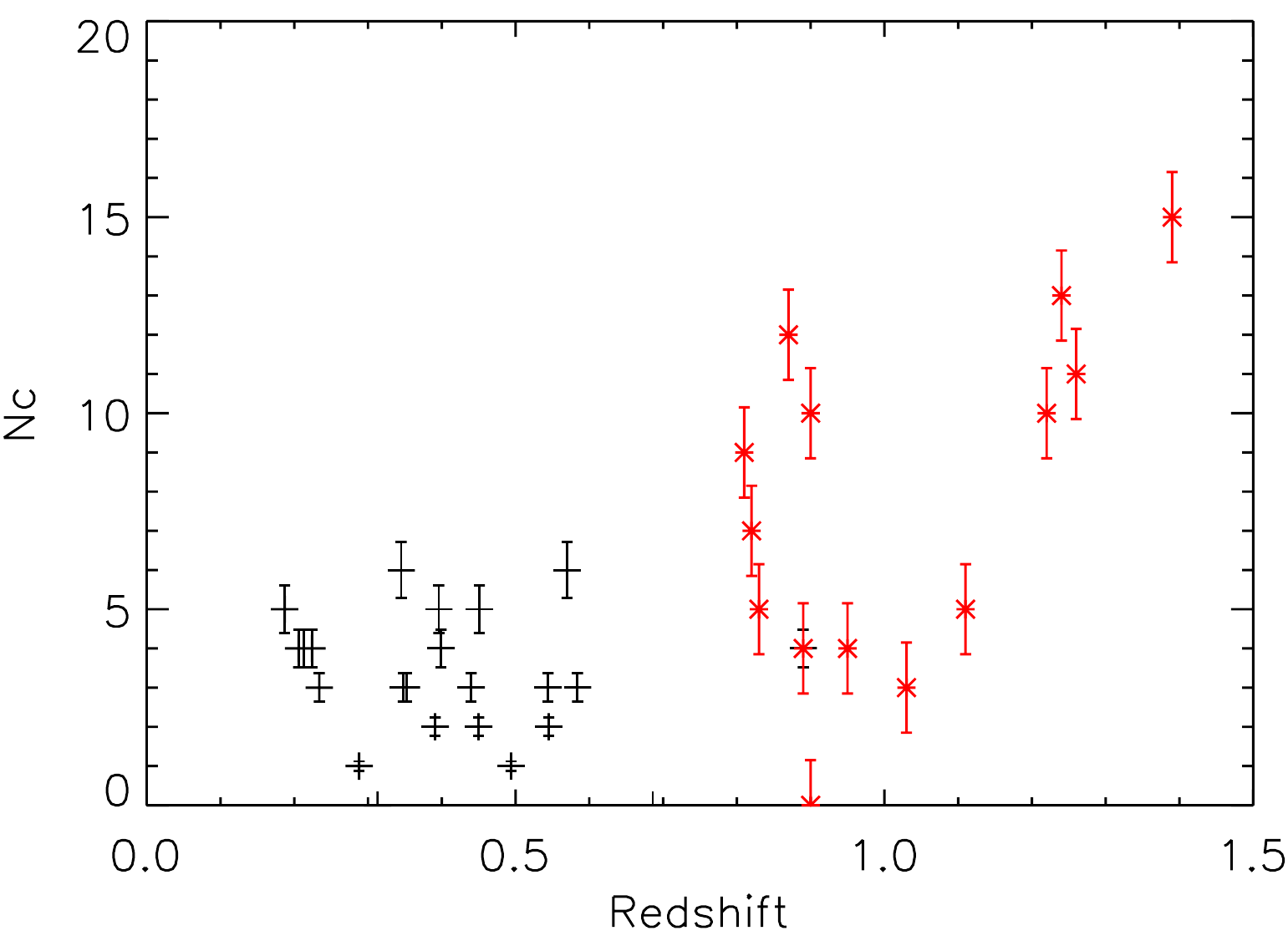}
\caption{The number of BCG companions (Nc) with redshift from CLASH (black) and Burke \& Collins 2013 (red crosses), where the detection limit for companions is consistent between the two samples. The errors on the CLASH points are smaller than those on the BC13 points due to the more accurate method by which non-cluster galaxies were excluded from the companion counts, see Section~\ref{analysis_section}.}
\label{nc_bc_comp}
\end{figure}

\begin{figure}
\centering
\includegraphics[width=8.5cm]{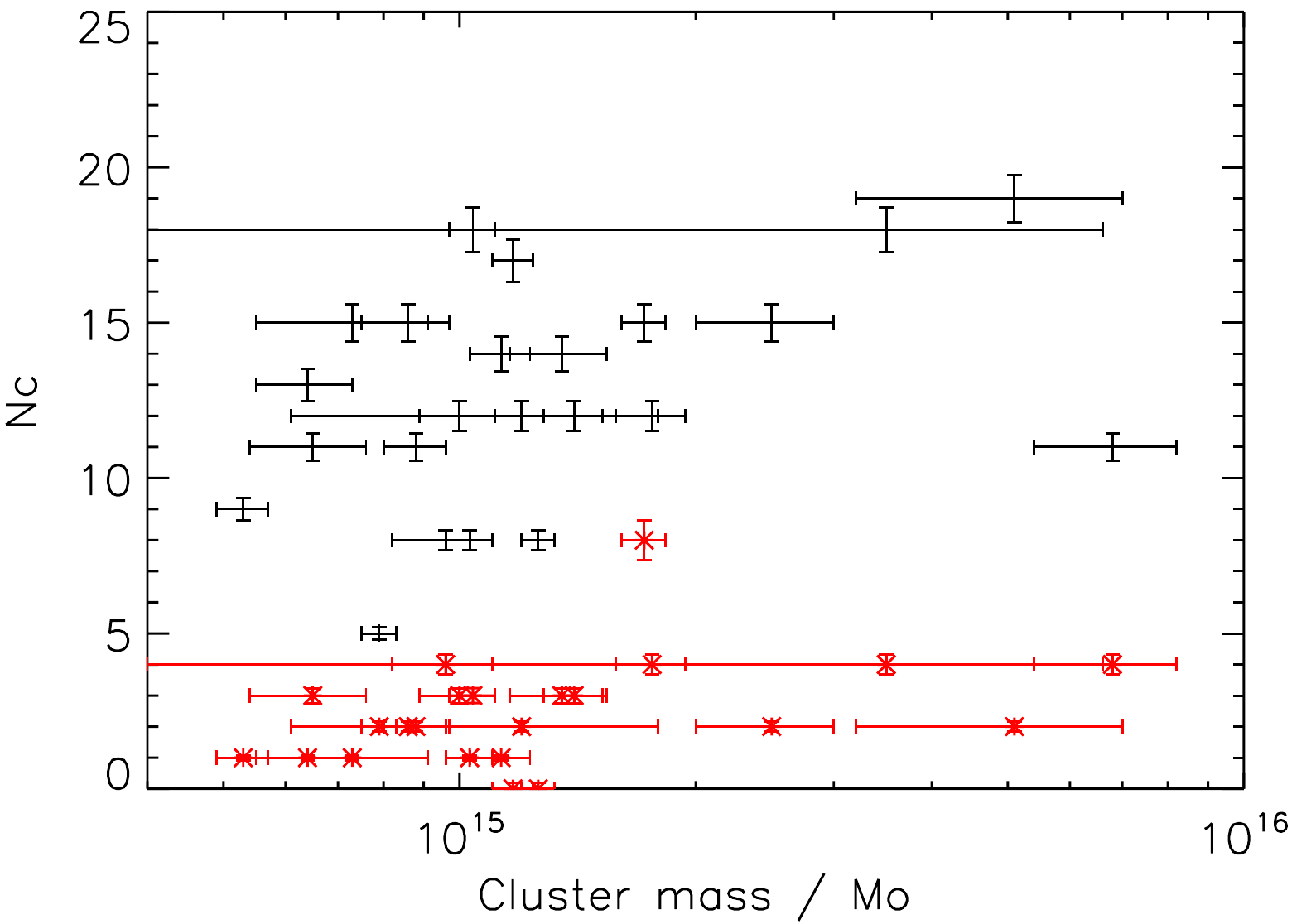}
\includegraphics[width=8.5cm]{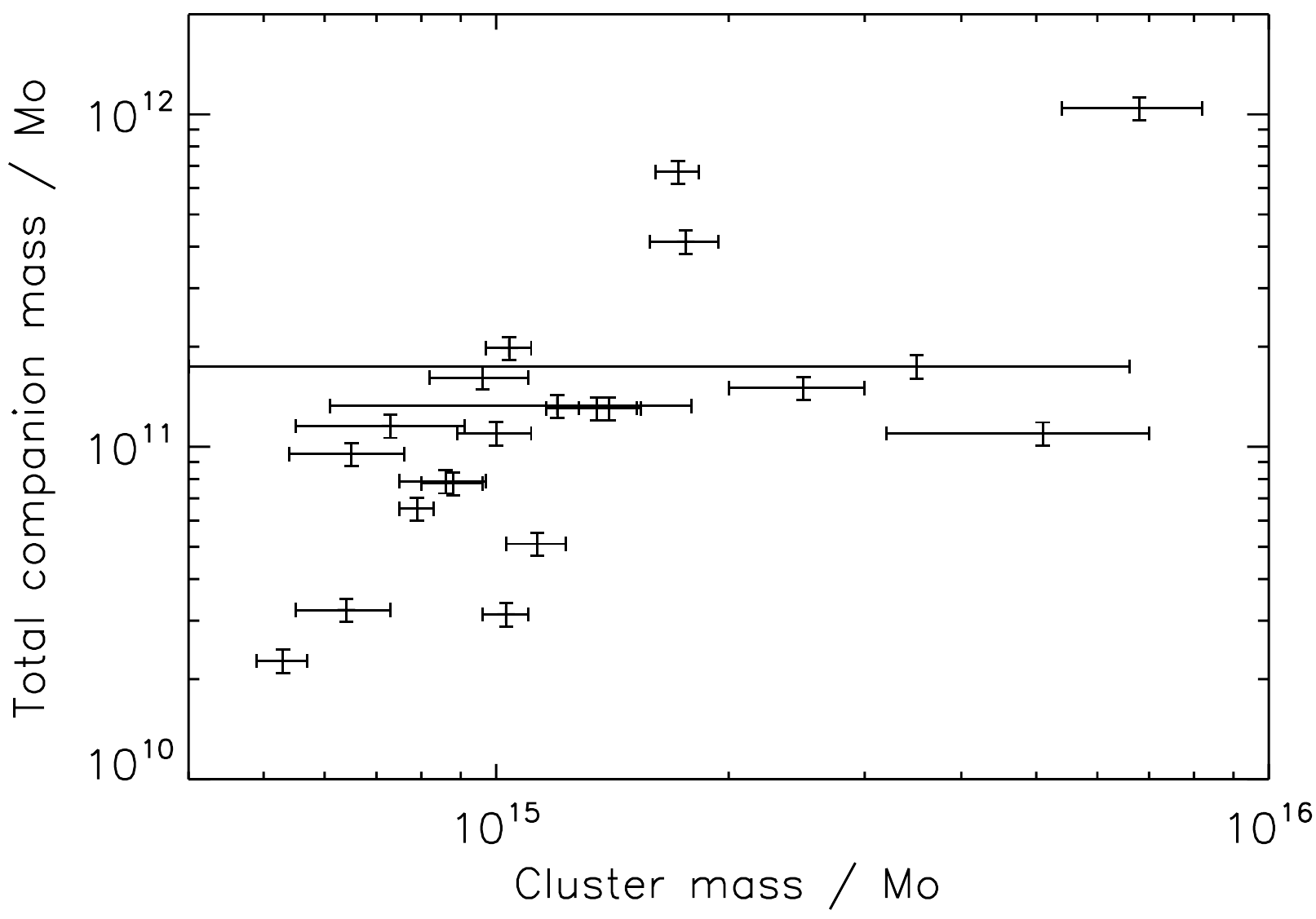}
\caption{Top: The relation between the number of BCG companions within 50 kpc and cluster mass ($M_{200}$). Black points show total companions, red crosses show companions which will merge with the BCG.
Bottom: Total mass of BCG companions which will merge with the BCG according to dynamical friction timescale, compared to the cluster total mass.}
\label{nc_cl_mass}
\label{comp_cl}
\end{figure}

\begin{figure}
\centering
\includegraphics[width=8.5cm]{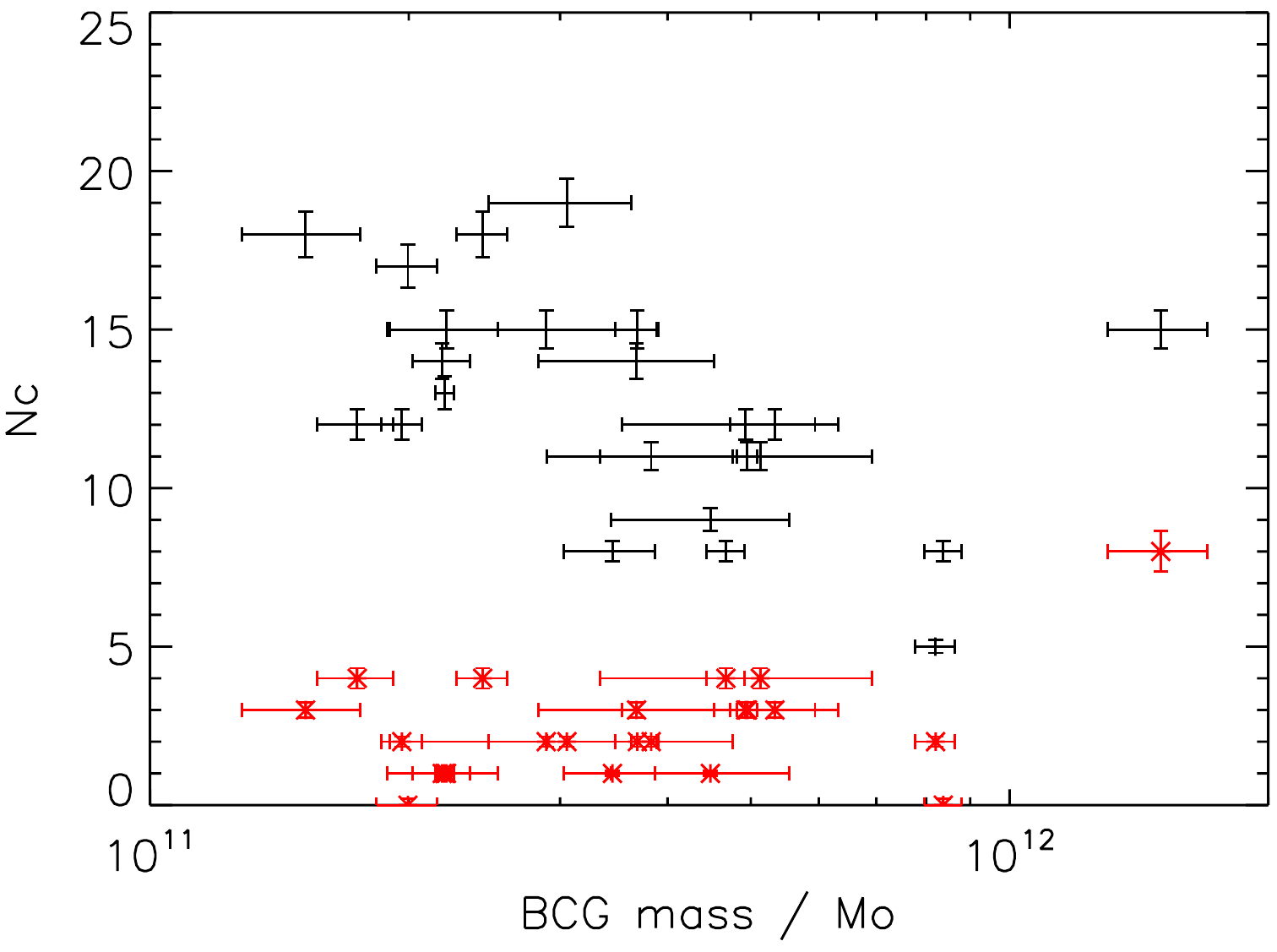}
\includegraphics[width=8.5cm]{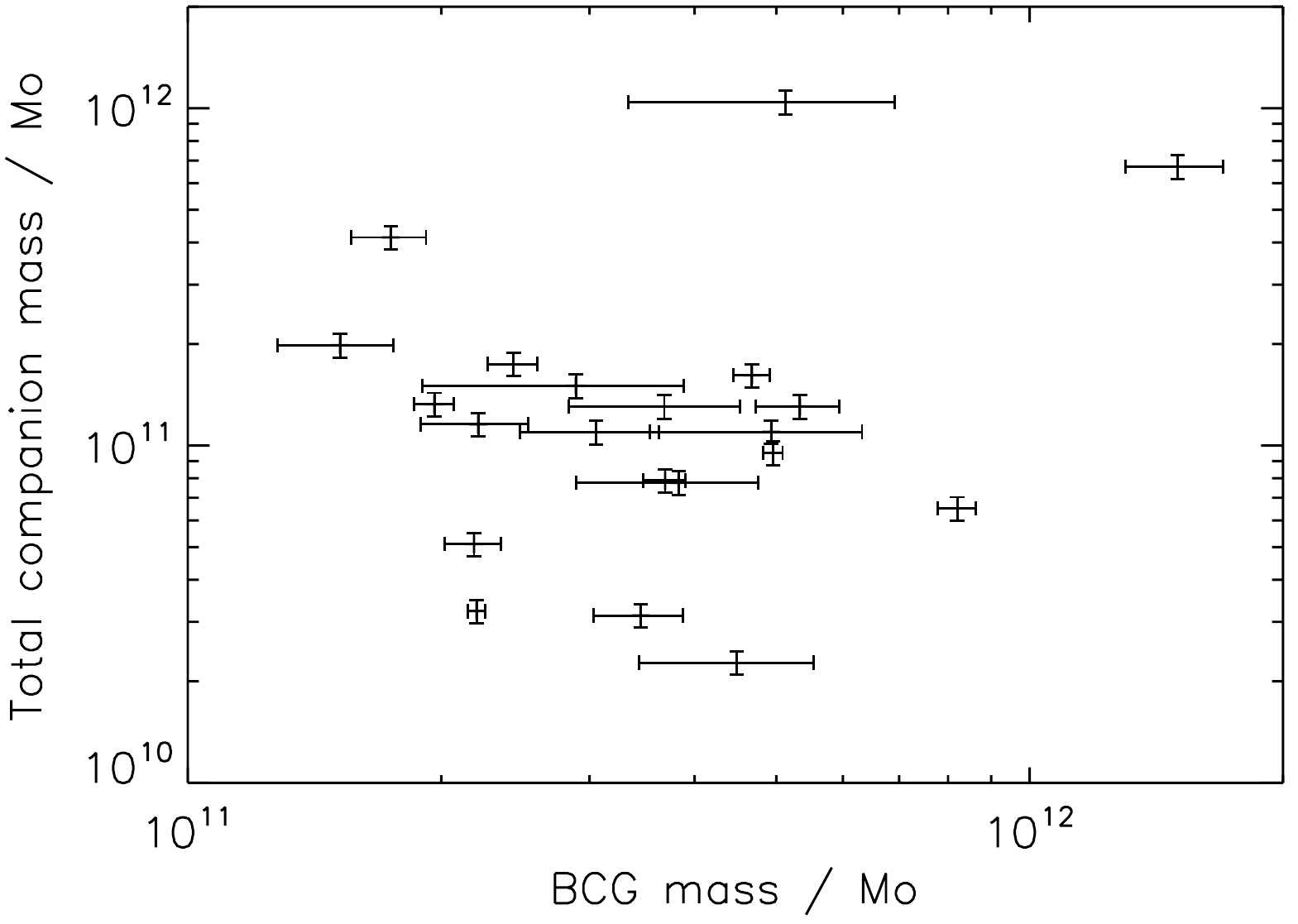}
\caption{Top: The number of BCG companions (Nc) within 50~kpc vs BCG mass for CLASH. Black points show total companions, red crosses show companions which will merge with the BCG. 
Bottom: Total mass of BCG companions within 50 kpc that will merge with the BCG according to dynamical friction timescale, compared to the BCG mass at its redshift.}
\label{bcg_nc}
\label{comp_bcg}
\end{figure}

Now we turn our attention to the numbers of BCG companions. The top panel of Figure~\ref{nc_z} shows the redshift relation of the total number of BCG companions within 50~kpc and the number of these companions for which their dynamical friction timescale indicates that they will merge with the BCG in the time between the cluster redshift and the present. Both the total $N_c$ and the merging $N_c$ are highly uncorrelated with redshift, with SR= $-$0.1 (SDZ= 0.6) and SR=0.45 (SDZ= 0.03) respectively. Figure~\ref{nc_bc_comp} shows the relation between $N_c$ and redshift when the number of BCG companions in CLASH is limited to the same magnitude completeness as that of BC13. This figure shows a decrease in the number of BCG companions with redshift, however this trend is fairly weak, with SR=0.5, but highly significant, SDZ=10$^{-3}$. In Figure~\ref{nc_bc_comp} all the companions have luminosities greater than 1:20 of the BCG, whereas Figure~\ref{nc_z} shows companions down to a luminosity ratio of 1:1000 of the BCG for the total $N_c$; thus this implies that for more massive BCG companions there is a general decrease in their numbers with redshift. This decrease is certainly to be expected for the most massive BCG companions as they have much shorter merging timescales than less massive companions, and this shows that there are several major and large-sized minor mergers (BCG:companion masses $\sim$ 1:3--1:10) in the cluster core between $z\lesssim1.5$ and the present. This also suggests that the larger total $N_c$ shown in the top panel of Figure~\ref{nc_z} is a result of the presence of many low luminosity companion galaxies, which are not detected in BC13. The presence of more low luminosity galaxies is corroborated in Figures~\ref{massg_frac} and \ref{nc_lit_comp} and is discussed below. It is worth noting that for the one overlapping cluster between the two samples, CLJ1226, the same number of companions is measured from the data used here as from BC13 when the same magnitude limit for companion detection is applied for both datasets.

The number of BCG companions shows no clear trend with cluster mass, for CLASH companions that both will and won't merge (top panel Figure~\ref{nc_cl_mass}, black points SR=0.4, SDZ= 0.08; red points SR=0.4, SDZ=0.05); clear trends are also not seen when the higher redshift BC13 sample is included (SR=$-$0.2, SDZ=0.2). The top panel of Figure~\ref{bcg_nc} shows the relation between the number of BCG companions within 50 kpc and BCG stellar mass. A possible weak trend is found in this plot (SR=$-$0.53 $-$0.53, SDZ= 10$^{-3}$) but no trend is seen for the number of companions which will merge (SR=0.22, SDZ=0.3). Given that CLASH represents a sample with a small range of BCG masses it is perhaps not surprising to see no trend in the numbers of merging BCG companions when compared to BCG masses.

The bottom panels of Figures~\ref{comp_z}, \ref{comp_cl} and \ref{comp_bcg} show the total stellar mass contained in the BCG companions compared to redshift, cluster total mass and BCG stellar mass respectively. The bottom panel of Figure~\ref{comp_z} shows both the total companion stellar masses and the stellar masses of the companions which will merge with the BCG by $z=0$. This figure shows that the majority of the total companion stellar mass, on average 68\% of the total, will merge with the BCG. However, in the top panel of this figure it is shown that only around 30\% of the total number of companions will merge. It seems that the merging companions are indeed more massive, and the mass contained in all the less massive companions is smaller despite their larger numbers. However, there is no correlation between redshift and total companion stellar mass both for all the companions (SR=0.1, SDZ= 0.7) and just those that will merge (SR=0.2, SDZ=0.4).

The bottom panels of Figures~\ref{comp_cl} and \ref{comp_bcg} show only the mass in the companions which will merge with the BCG by $z=0$.  Figure~\ref{comp_bcg} shows no correlation between BCG mass and merging companion mass (SR=$-$0.05, SDZ=0.8) and Figure~\ref{comp_cl} shows a weak, positive correlation between total cluster mass and merging companion mass (SR=0.6, SDZ= 0.01) suggesting the possibility that more massive companions reside in more massive halos. In a recent study by \citet{Tal_14} of massive galaxies in groups at 0.2$<z<1.2$ it was found that in more massive haloes, more of the total group mass was in the satellites of the central galaxy. It was also found that the companion masses are not correlated with redshift for fixed central galaxy mass, consistent with what is seen here. \citet{Tal_14} find that more massive central galaxies have more massive companions which is not seen here, however CLASH shows no correlation between BCG mass and halo mass - a trend which is usually observed - so the absence of this trend here is most likely due to the small dynamic range of BCG masses.

When we examine the relation between the number of companions to the BCG and the fraction of the total cluster light contained in the ICL (for both the total $N_c$ and just those that will merge) we find a large scatter and no obvious trends. The total $N_c$ has no correlation  with ICL fraction with SR=0.01 (SDZ=0.9), and the numbers of merging companions also show very little evidence of correlation with the ICL fraction (SR=$-$0.4, SDZ=0.03).

The merging companion masses also show no trend when compared to the ICL fractions for each cluster (SR=$-$0.08, SDZ=0.7). When the BCG mass and companion masses are summed together there is also very little evidence for correlation ( SR=$-$0.6, SDZ=0.02). This lack of correlation suggests that the numbers of companions and their masses has at best a weak effect on the amount of ICL at a given redshift.  However the number of companions at a given redshift does not provide any information about the number of mergers that the cluster core may have experienced, hence this gives no insight into the effect of mergers on the growth of the ICL fraction.

\begin{figure}
\centering
\includegraphics[width=8.5cm]{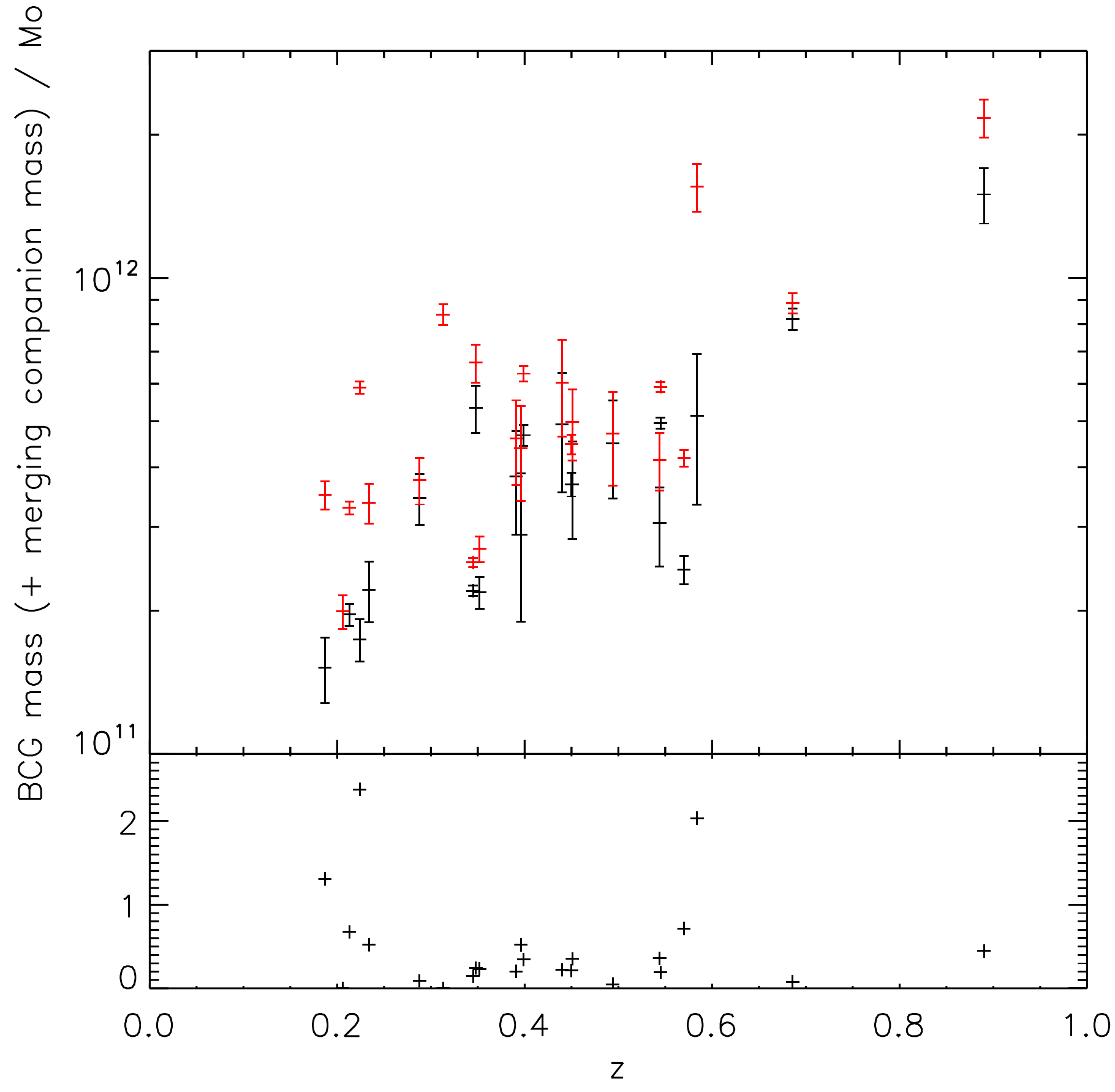}
\caption{The inferred mass growth additional to the BCG mass measured at the cluster redshift from the accretion of its companions within a 50~kpc aperture. Top: black points show the measured BCG mass at its current redshift, red points show the BCG mass after the accretion of its companions. Bottom: the mass growth of the BCG from its accreted companions divided by the BCG's original mass.}
\label{massg_z}
\end{figure}

\begin{figure}
\centering
\includegraphics[width=8.5cm]{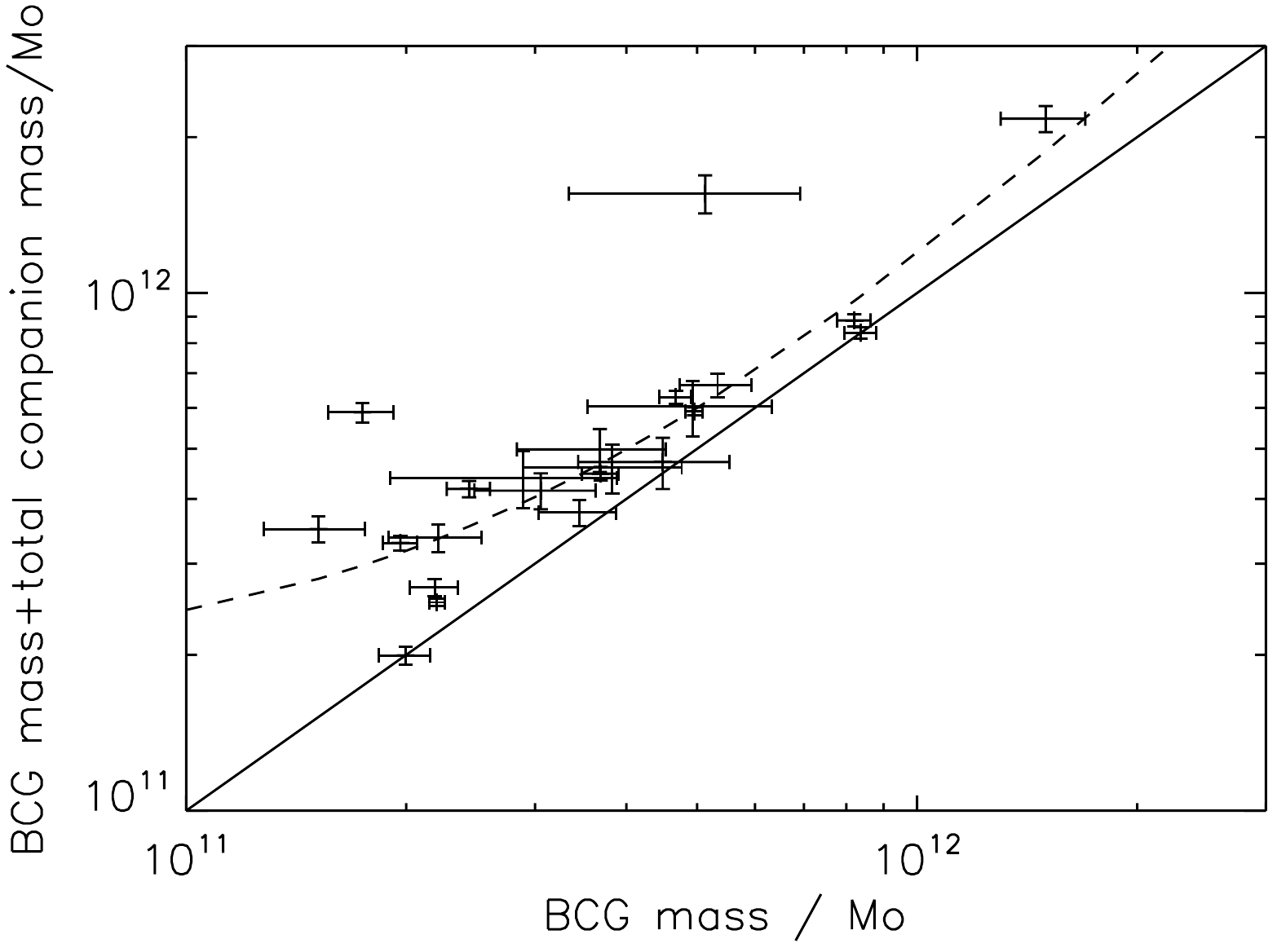}
\caption{BCG mass vs the BCG mass after merging of companions with dynamical friction timescales less than the time between $z=z$(BCG) and $z=0$. The solid line represents x=y, the dashed line is the best fit for a power-law fit.}
\label{bcg_massg}
\end{figure}

Figure~\ref{massg_z} shows the BCG masses after the accretion of their companions as a function of redshift, with the original BCG mass for comparison. The BCG masses tend to be similar before and after the addition of the merged companion masses in most cases. For a linear fit the gradients for both sets of points are within 2$\sigma$ of each other, showing no redshift dependance on final BCG mass. The lower panel of Figure~\ref{massg_z} shows the BCG mass growth divided by the original BCG mass against redshift, and this also shows no obvious trend in mass growth with redshift. For 11 of the 23 clusters the BCG mass after accretion of companions is the same within the errors as the BCG mass at the cluster redshift. Figure~\ref{bcg_massg} shows this more clearly, where the BCG measured mass is plotted against the BCG mass after accretion of its companions.  The solid line on this figure indicates $y=x$ and the dashed line is a geometric $y=ax^b+c$ best fit, for which a=(3$\pm$2)$\times10^{-4}$, b=1.29$\pm$0.2, c=(1.93$\pm$0.3)$\times10^{11}$. The possible slight increase in the steepness of the curve appears to be from the growth of the least massive BCGs; however both the change of the slope and the mass growth of the BCGs are moderate at best. The maximum stellar mass growth for the whole sample is 3.4 times the original BCG mass, but the mean growth is more moderate at 1.4 times. 
The amount of accreted matter from a merger which ends up falling on to the central BCG or is instead distributed within the diffuse ICL is observationally undefined. Simulations of the buildup of the ICL  predict that the fraction of the stellar matter which goes into ICL rather than BCG during a merger is between 30--80\% \citep{Murante_07, Laporte_13, Conroy_07, Contini_14, Puchwein_10}. Assuming an average of 50\% of merging stellar mass goes into ICL rather than BCG, the subsequent BCG mass growth for the CLASH sample should be 1.2 times the original BCG mass. 
This suggests that on average BCGs do not grow by a large amount in stellar mass from the accretion of their companions. 

The BCG after accretion of its companions shows no correlation with cluster mass (SR=0.2, SDZ= 0.4). The BCG growth also shows no significant correlation with redshift or original BCG mass (as shown in Figures~\ref{massg_z} and \ref{bcg_massg}) and we find only moderate growth in stellar mass. However, the ICL shows no significant increase over the same redshift range and it is apparent that the growth of the ICL is the major evolution event in the cluster core during the second half of the lifetime of the Universe.

\begin{figure}
\centering
\includegraphics[width=8.5cm]{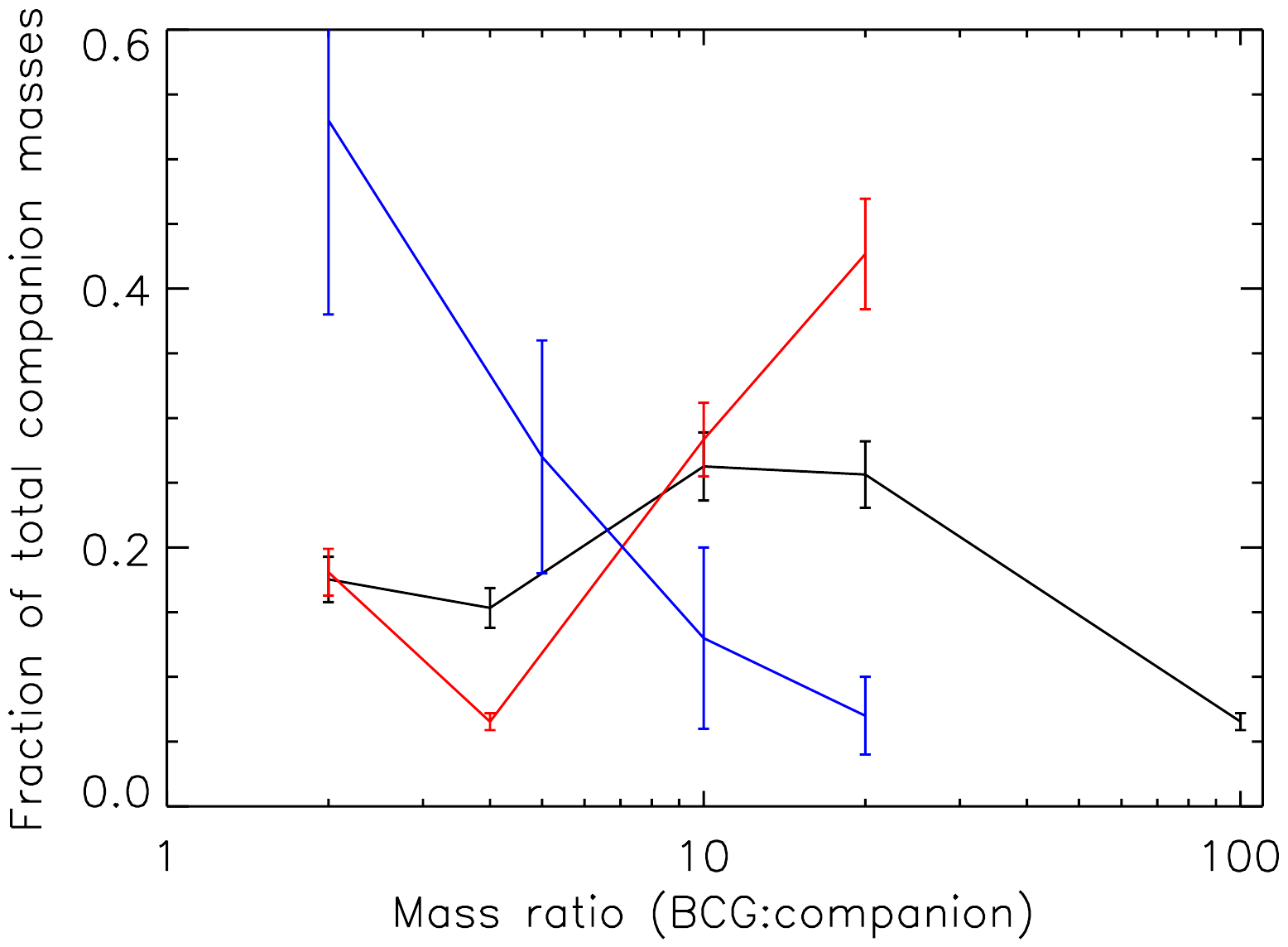}
\caption{The fraction of the BCG mass growth from the accretion of its companions between the cluster redshift and the present compared to the mass of the companion - companion mass decreases from left to right. Black: total merging companions in CLASH; red: CLASH companions limited to the same luminosity as the completeness of \citet{BC13}; blue: results of \citet{BC13}.}
\label{massg_frac}
\end{figure}

Figure~\ref{massg_frac} shows the contribution to the stellar mass growth of the BCGs that accreted companions will make as a function of their stellar mass ratio with respect to the BCG mass. The CLASH results are shown for companions which will be accreted (black) and for companions as limited to the same completeness as BC13 (red), alongside the results for BC13 (blue). This figure shows that $\sim$25\% of the total merging mass is in companions of mass ratio to the BCG of 1:5--1:10 and 1:10--1:20. A further $\sim$16\% comes from companions of mass ratios of both 1:2--1:3 and 1:3--1:5 with the BCG. Finally only $\sim$6\% of the merging companion mass is in galaxies of 1:20--1:100 of the mass of the BCG. Added together the majority of the merging mass in BCG companions, around 60\% is in low mass BCG companions of masses less than 1/5 of the BCG, showing that for CLASH the majority of mergers in the cluster core will be minor mergers and there will be very few major mergers. When compared with the results of BC13 (for the points which are limited to the same companion brightness completeness limit) it is clear that there is an abundance of low mass companions in the CLASH cluster cores compared to the sample used in BC13. This is to be expected as the BC13 sample is at a higher average redshift than CLASH (c.f. $z=1$ for BC13 and $z=0.4$ for CLASH) and the more massive companions which have shorter dynamical friction timescales will have merged in the cluster core between the average sample redshifts of $z=1$ and $z=0.4$. This indicates that BCGs at higher redshifts are built by major mergers (or that major mergers are the dominant source of stellar mass assembly in the cluster core), and minor mergers are the more dominant merging path at lower redshifts.

\begin{figure}
\centering
\includegraphics[width=8.5cm]{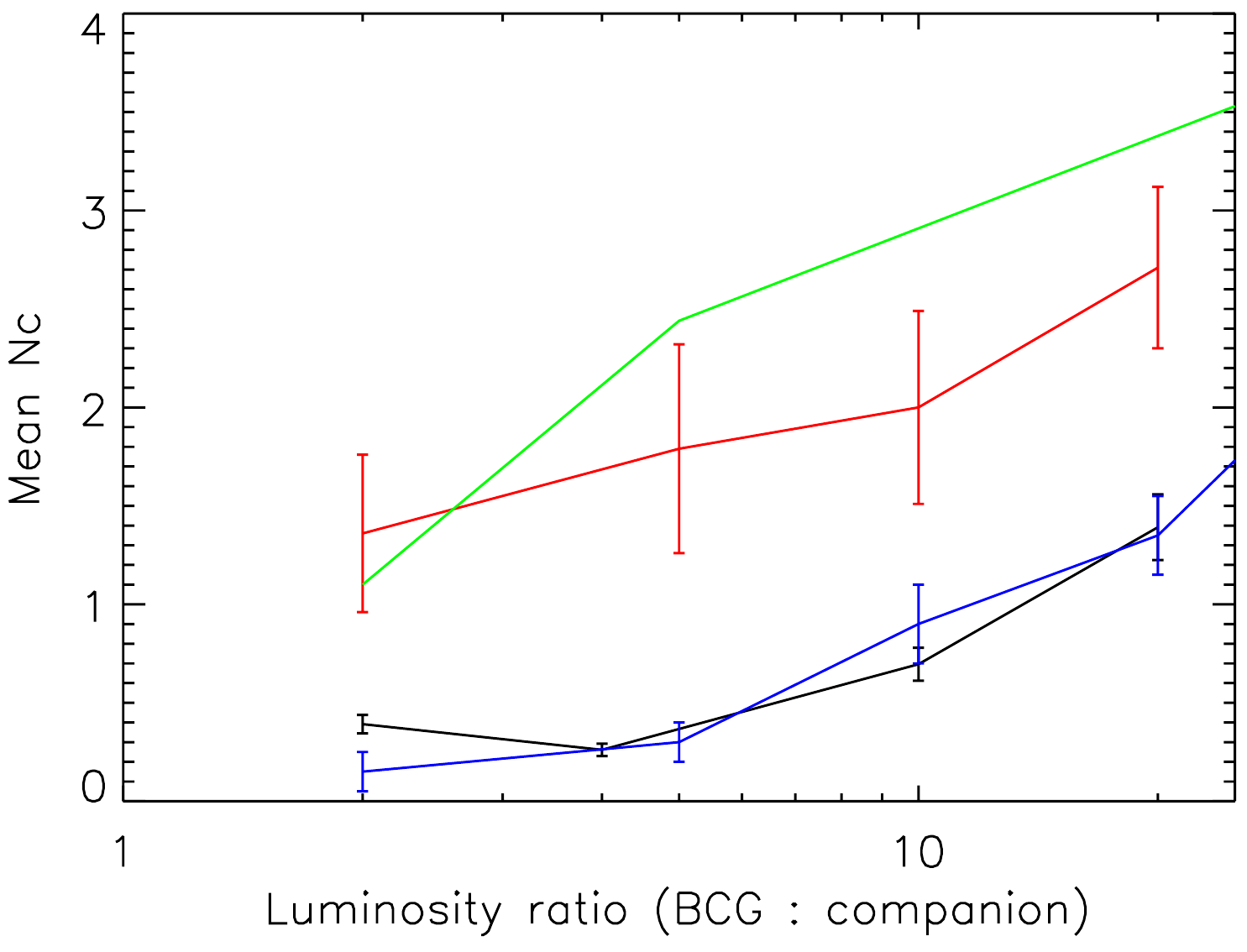}
\caption{The numbers of BCG companions split by luminosity ratio to the BCG compared to results from the literature. Black: this study, red: BC13, blue: \citet[$z\sim0.3$]{Edwards_12}, green: simulations of \citet[$z=$2--0]{Laporte_13}.}
\label{nc_lit_comp}
\end{figure}

When the companions are split into luminosity ratio (and hence mass ratio) bins, as would be expected there are fewer BCG companions with masses comparable to the BCG, and there are very many small mass companions at all redshifts. Figure~\ref{nc_lit_comp} shows the mean numbers of BCG companions by luminosity ratio bin compared to results from previous studies. The CLASH clusters show numbers of companions comparable to (the same within the errors) those found by \citet{Edwards_12}, who examined a sample of clusters at $z\sim$0.3. The mean redshift of the CLASH sample is $z=0.4$, so it is not surprising to find good agreement between these results. The other two studies included in this figure are the observational study of \citet{BC13} and the simulated predictions of \citet{Laporte_13}, both of which examine BCGs at higher redshifts, $0.8<z<1.2$ and $z\sim$2 respectively. Since less time has passed for the higher redshift clusters it is not surprising that these higher redshift studies show and predict a larger number of companions in all mass ratio bins, and in agreement with the results shown in Figure~\ref{nc_bc_comp} there is an overall trend between redshift and the total number of BCG companions between $z\sim1.5$ and $z=0$, which is more obvious when the BCG companions are split by their comparative luminosity ratio to the BCG.

\section{Discussion}\label{discussion}

Across the CLASH sample the stellar masses of the BCGs and the total cluster masses both cover a fairly narrow mass range. This is also reflected in the lack of correlation between BCG stellar mass and cluster mass, where one would normally expect to see a positive correlation between the two. This narrow mass range of both clusters and BCG stars is likely to be a selection effect. However, given this fairly consistent mass sample, we find strong evidence for the substantial growth of the ICL and evidence of mergers which would lead to a moderate growth of BCG stellar mass or alternatively an increase in stars in the ICL.

Typically galaxy clusters which exhibit strong lensing features are extended along the line of sight. If this were the case we would expect to measure higher than `normal' ICL fractions and numbers of BCG companions for the strong lensing CLASH clusters. The selection criteria as described in \citet{Postman_12} was specifically designed to avoid this, with clusters selected for their spherical symmetry in X-rays. However if these clusters are highly relaxed as their X-ray symmetry would suggest then it may be that we should expect them to be highly evolved, having already undergone the majority of their galaxy merging interactions thus showing high BCG masses, large ICL fractions and low numbers of BCG companions or only low-mass BCG companions. Eight of the clusters in the sample are reported to show some small degree of substructure (A209, MACS0329, CLJ1226, MACS0744, MACS1206, RXJ1347, A2261, RXJ2248), making them not fully relaxed. These eight are spread across the whole redshift range of CLASH and do not show any noticeable deviation from the general trends seen in the results, suggesting that selection criteria is not a major issue for the results from this study.

According to the dynamical friction timescale for mergers of BCG companions, the BCG stellar masses only increase moderately over the time between 0.18$\le z\le$0.9, on average growing by a factor of 1.4. The eventual resting place of accreted matter in cluster cores is unknown, and when we assume 50\% of merging mass ends up in ICL rather than centrally on the BCG, in line with the predictions of simulations, the mass growth of the BCGs is only a factor of 1.2 on average. Given the observed large growth of the ICL over the same time, it seems likely that the overwhelming majority of companion mass must end up in ICL rather than BCG. 

The growth factor for BCGs from mergers of 1.4 (1.2 if we assume 50\% of accreted matter ends up in the ICL) reported here is fairly consistent with the stellar mass growth estimate of 1.8 for BCGs in the time between $z =$ 0.9--0.2 reported by \citet{Lidman_12}. \citet{Collins_09} find $>$90\% of BCG stellar masses in place by $z =$ 1; and \citet{Stott_10} find BCGs at $z =$ 1 on average have 95\% of the stellar mass of those measured locally, which corresponds to growth of factors of 1.1 and $>$ 1.05 times respectively. This is also fairly consistent with the factor of 1.4 or 1.2 found here. 
\citet{Laporte_13} simulate BCG mass growth by mergers and predict BCG stellar mass growth factors of 1.5 between $0 < z < 0.3$, 1.9 between $0.3 < z < 1$ and 2.6 between $0 < z < 1$. They also predict that $\sim$30\% of accreting mass goes into the ICL rather than the BCG. The simulation of \citet{DeLucia_07} predict BCG stellar mass growth of a factor of 4 since $z\sim$1 and a factor of 1.6 since $z\sim$0.4, however \citet{DeLucia_07} have no prescription for the ICL in their simulation. The implied mass growth we find for CLASH lies in between the previous observational results  of Collins et al., Stott et al. and Lidman et al., but is notably less than the growth predicted by simulations.

The stellar population of the ICL is largely unknown, and if it is built from the accretion of other cluster galaxies then its stellar population may be quite varied from location to location within a cluster. Available estimates use the colour of the ICL to derive its stellar population \citep[e.g.,][]{Krick_07} and suggest that the ICL near the cluster core has a population similar to that of BCGs, being old and red (however \citealt{Krick_07} also find gradients in their ICL colours). If we assume that the stellar population of the ICL is the same as that of the BCG we can give a rough estimate of its mass and its mass growth over the redshift range of this study. Figure~\ref{bcg_icl_frac} shows the relative contributions of the BCG and ICL to the total cluster light. If we assume that the stellar mass scales directly with stellar luminosity for the same kind of stellar population, this figure would suggest that the ICL has 5--10 times more stellar mass than the BCG at low redshifts $z<0.4$, and a similar mass to the BCG at higher redshifts $z\sim0.4$. The average BCG mass for CLASH is $\sim 4\times10^{11}M_{\odot}$. Taking a conservative value of 5 times the mass of the BCG gives an average estimate for the ICL stellar mass at $z<0.4$ of 2$\times10^{12}M_{\odot}$. Given this estimate for the ICL mass at low redshift, the ICL must accrete approximately 1.6$\times10^{12}M_{\odot}$ of stellar mass, presumably from merging companions since $z\sim0.4$. This is a significantly higher mass than is contained in all the companions to the BCG within 50~kpc for all the CLASH clusters, by around a factor of 5--10. If the mass in the ICL is indeed this high, and it grows substantially via tidal stripping from mergers, then the majority of the mass which grows the ICL must come from outside of 50~kpc, as is predicted by some simulations of ICL growth \citep{Conroy_07, Murante_07, Purcell_07, Puchwein_10}.

An amount of stellar mass accretion of the magnitude estimated above adds up to a lot of mergers, especially if the amount of accreted matter that ends up deposited on the ICL is only 50\% of the total merging mass. This suggests that our assumed value of 50\% may be a very conservative estimate, and that the actual fraction of accreted matter from mergers which contributes to the ICL may be significantly higher.

We find that the fraction of the total cluster light which is contained within the ICL grows significantly, by a factor of 4--5 since $z=0.4$. This result is broadly in line with previous observational studies and is consistent with the predictions of simulations which make special consideration for the ICL,  however the rate at which the ICL grows is somewhat different to that predicted by simulations. Given that simulations generally predict the ICL to assemble through galaxy mergers and interactions with the BCG (including tidal stripping), it is clear that the consideration of the ICL is vital when examining the growth and assembly of BCGs, especially in light of the, at most, moderate mass growth observed for BCGs.

 We find that the amount of ICL detected is not only dependant on the upper limit of the surface brightness threshold below which it is measured, but it is also strongly dependant on the faint-end surface brightness limit.  As is shown in Figure~\ref{bcg_icl_frac} we find a 40\% difference in the fraction of ICL recovered when we change the faint surface brightness limit by 0.5 mag/arcsec$^2$. The faint-end, lower surface brightness limit above which flux is measured is generally not discussed in previous studies of ICL, both observational and simulated. To illustrate this, \citet{Presotto_14} examine the ICL in one of the clusters in CLASH, MACSJ1206 ($z=0.44$), using data from the Subaru Suprime-Cam. They measure the ICL to contain 12.5$\pm$0.6\% of the total cluster light below a surface brightness threshold of $\mu_V=$26.5 mag/arcsec$^2$. The \citet{Presotto_14} data is deeper than the CLASH HST data used here and therefore can measure ICL to a rest-frame surface brightness $\mu_V=$28.5 mag/arcsec$^2$, significantly deeper than is possible here. 
In order to compare more accurately between different observational studies and between observations and simulations of the ICL the effect of the surface brightness lower limit needs to be more fully quantified.

When we examine the respective contributions of the BCG and ICL to the total cluster light we find that  the ICL contains 70--80\% of BCG+ICL light  at redshifts $z<0.2-0.3$. The fraction of BCG+ICL light in the ICL shows a significant increase as redshift decreases. This result also points to the importance of the ICL as a component within cluster cores and again shows the significant growth of the ICL in contrast to the only moderate growth of the BCG. The ICL having more of the total cluster light than the BCG is also observed in previous studies \citep[e.g.][]{Gonzalez_07, Gonzalez_13}, however these studies use lower cluster-mass and redshift samples than CLASH. The CLASH clusters on their own do not show the negative correlation between BCG+ICL light as a fraction of the total cluster light and cluster total mass as is observed in these previous studies (see Figure~\ref{bcg+icl}), however this may be due to the smaller dynamic range of cluster masses in CLASH or the lower redshifts of the samples used in other studies.

When comparing the numbers of companions between studies across a range of different redshifts we see fairly strong evidence for the occurrence of mergers in cluster cores. The BCG companions within a 50~kpc radius are generally less massive at lower redshift and there are generally fewer of them. Similar to the comparable redshift study of \citet{Edwards_12}, we find a greater number of small-mass companions (mass ratios 20:1) than large-mass companions, and find very large companions (masses $\ge$2:1 w.r.t. BCG) to be rare. However at the higher redshifts probed by \citet{BC13}, significantly more large mass companions are seen. This suggests that major mergers are an important mechanism for stellar mass build up at high redshift and minor mergers are more dominant at low redshift. Given the lack of BCG stellar mass growth coincident with a large ICL growth it seems that minor mergers build up ICL at medium to low redshift. 

\citet{Edwards_12} estimate a factor of 1.1 growth in BCG stellar mass from the accretion of their measured companions since their average redshift $z=0.3$ (10\% increase in mass), which is similar to the value of 1.4 (1.2) found here, however \citet{Edwards_12} measure much lower numbers of companions compared to that found for the CLASH sample, with their sample having 0.7 companions per BCG on average, c.f. CLASH has 12.7 companions per BCG on average, of which 2.4 will merge according to dynamical friction timescales. Other observational studies of low redshift BCGs also find small numbers of companions and major merging to be rare. For example \citet{Liu_09} use a sample 515 BCGs at z $<$ 0.2 and find large companions (with luminosity ratios less than 1:4) within 30 kpc of the BCG to be rare, with an average 0.1 of these companions per BCG (c.f. 0.7 for CLASH). Liu et al. find 18/515 of their BCGs to be undergoing major mergers and suggest that BCGs should have increased their masses by a factor of 1.15 (15\%) from major dry mergers since z = 0.7. \citet{McIntosh_08} find only 38/845 BCGs at z$\leq$0.12 to be undergoing major mergers (0.04 mergers per BCG) and predict that massive haloes are growing by 1--9\% per Gyr by major mergers (factor $\le$1.09 per Gyr). Whilst these studies predict similar mass growths to that found for the CLASH sample, we do not see any major mergers currently taking place in CLASH (this is likely a result of the selection of the sample for relaxed-looking clusters) and find that minor mergers are far more likely at these low redshifts. 
The higher-redshift sample presented in BC13 generally finds more larger BCG companions, suggesting that major merging is more important for the stellar mass build-up of high-redshift BCGs than nearby BCGs. For companions more massive than 1:5 of the BCG, BC13 find 1.3 companions per BCG, twice as many as the 0.7 found for CLASH, and as is evident in Figure~\ref{nc_lit_comp}  BC13 find more than twice as many companions in all mass bins compared to CLASH, showing that the majority of companions of all masses should indeed have merged with the BCG in the time between $z\sim$1 and $z\sim0.4$.

\section{Conclusions}\label{conc_s}
We have studied the BCG stellar mass growth, ICL growth and numbers of mergers onto BCGs for a sample of massive clusters across a redshift range spanning the second half of the lifetime of the Universe. We have found the clusters in the sample to have a narrow range of cluster and BCG masses, and that the BCG masses should only grow by a small amount from mergers of the surrounding companions, on average a factor of 1.4, or a factor of 1.2 when the  amount of merging mass which simulations predict will end up in the ICL is considered. Conversely we find that the ICL grows by a substantial amount over the same timespan and for the same clusters, increasing its contribution to the total cluster light by a factor of 4--5 over this time.  We also suggest that the growth of stellar mass in the ICL is larger than can be provided be the BCG companions within 50~kpc, and that the majority of the ICL mass must come from galaxies which  fall from outside of the core of the cluster.  One caveat to quantifying the ICL growth is the depth at which observational data is taken and consistency in comparison between ICL surface brightness limits used in model predictions.

We conclude that BCGs must have assembled the majority of their stellar masses very early on in the life of the Universe, before a redshift of $z=1$. We suggest that the majority of stellar mass growth of BCGs at high redshifts ($z\gtrsim 1$) is by major merging, and the majority of merging activity in the latter half of the lifetime of the Universe is by stripping and minor merging in which the majority of the accreted or disrupted stellar mass ends up contributing to the ICL rather than the BCG.

\section*{Acknowledgments}
We acknowledge Ivan Baldry for assisting us with the K-correction. We thank the anonymous referee for their comments which greatly improved the quality of this paper. CB and MH acknowledge support from the National Research Foundation Multi-Wavelength Astronomy Programme and the University of KwaZulu-Natal. CAC acknowledges support from STFC Consolidated Grant  ST/J001465/1. 

\bibliographystyle{mn2e}
\bibliography{refs}

\pagestyle{empty}
\begin{landscape}
\begin{table}
\begin{center}
 \label{results_tab}
\caption{Results for CLASH sample BCG stellar masses, fraction of the total cluster light contained in the ICL, total number of companions to BCG within 50~kpc and the number of these which will merge with the BCG according to their dynamical friction timescale, the masses of the merging companions and the merging companion mass compered to the BCG mass.}
 \label{results_tab}
\begin{tabular}{l c c c  c  ccccccccccc}
\hline
Cluster &  Redshift &  Cluster mass $M_{200}$ &  BCG mass &  \multicolumn{2}{c}{\% of total cluster light} &    $N_c$ total & $N_c$ merge&  Merging companion & Total companion mass   \\
 &  & ($\times 10^{15}M_{\odot}$) & ($\times 10^{11} M_{\odot}$) &  ICL &  BCG &  & &  mass ($\times 10^{11} M_{\odot}$) & /BCG mass    \\
\hline
Abell 383 & 0.187 & 1.04$\pm$0.07 & 1.51$\pm$0.24 & 23.05$\pm$0.73 & 2.89$\pm$1.19 & 20.0$\pm$0.8 & 3.0$\pm$0.2 & 1.98$\pm$0.16 & 1.31$\pm$0.13 \\	
Abell 209 & 0.206 & 1.17$\pm$0.07 & 2.0$\pm$0.16 & 17.11$\pm$0.77 & 2.86$\pm$1.37 & 16.0$\pm$0.6 & 0.0$\pm$0.2 & 0$\pm$0 & 0 \\	
Abell 1423 & 0.213 & 1.20$\pm$0.59 & 1.96$\pm$0.11 & 17.97$\pm$0.73 & 2.96$\pm$1.39 & 9.0$\pm$0.4 & 2.0$\pm$0.2 & 1.33$\pm$0.11 & 0.68$\pm$0.05 \\	
Abell 2261 & 0.224 & 1.76$\pm$0.18 & 1.74$\pm$0.18 & 16.64$\pm$0.78 & 1.78$\pm$1.16 & 11.0$\pm$0.4 & 4.0$\pm$0.3 & 4.14$\pm$0.33 & 2.38$\pm$0.15 \\	
RXJ2129+0005 & 0.234 & 0.73$\pm$0.18 & 2.21$\pm$0.32 & 12.89$\pm$1.11 & 3.04$\pm$1.43 & 16.0$\pm$0.6 & 1.0$\pm$0.1 & 1.16$\pm$0.09 & 0.52$\pm$0.09 \\	
Abell 611 & 0.288 & 1.03$\pm$0.07 & 3.45$\pm$0.42 & 13.02$\pm$0.23 & 3.49$\pm$1.05 & 8.0$\pm$0.3 & 1.0$\pm$0.1 & 0.31$\pm$0.03 & 0.09$\pm$0.06 \\	
MS 2137-2353 & 0.313 & 1.26$\pm$0.06 & 8.38$\pm$0.42 & 7.58$\pm$1.65 & 1.05$\pm$0.98 & 8.0$\pm$0.3 & 0.0$\pm$0.2 & 0$\pm$0 & 0 \\	
RXJ1532+30 & 0.345 & 0.64$\pm$0.09 & 2.20$\pm$0.05 & 7.44$\pm$0.17 & 2.68$\pm$1.13 & 13.0$\pm$0.5 & 1.0$\pm$0.1 & 0.32$\pm$0.03 & 0.15$\pm$0.02 \\	
RXJ2248-4431 & 0.348 & 1.40$\pm$0.12 & 5.33$\pm$0.61 & 6.44$\pm$0.10 & 2.42$\pm$0.63 & 12.0$\pm$0.5 & 3.0$\pm$0.2 & 1.31$\pm$0.10 & 0.24$\pm$0.07 \\	
MACSJ1115+01 & 0.352 & 1.13$\pm$0.10 & 2.19$\pm$0.17 & 6.11$\pm$0.29 & 1.39$\pm$1.09 & 14.0$\pm$0.6 & 1.0$\pm$0.1 & 0.51$\pm$0.04 & 0.23$\pm$0.05 \\	
MACSJ1720+35 & 0.391 & 0.88$\pm$0.08 & 3.83$\pm$0.94 & 3.16$\pm$0.06 & 1.49$\pm$0.31 & 11.0$\pm$0.4 & 2.0$\pm$0.2 & 0.78$\pm$0.06 & 0.20$\pm$0.13 \\	
MACSJ0416-24 & 0.396 & 2.50$\pm$0.50 & 2.89$\pm$0.99 & 2.69$\pm$0.10 & 1.08$\pm$0.10 & 15.0$\pm$0.6 & 2.0$\pm$0.2 & 1.50$\pm$0.12 & 0.52$\pm$0.19 \\	
MACSJ0429-02 & 0.399 & 0.96$\pm$0.14 & 4.68$\pm$0.23 & 2.36$\pm$0.13 & 1.21$\pm$0.59 & 8.0$\pm$0.3 & 4.0$\pm$0.3 & 1.61$\pm$0.13 & 0.35$\pm$0.04 \\	
MACSJ1206-08 & 0.440 & 1.00$\pm$0.11 & 4.93$\pm$1.39 &  & 1.21$\pm$0.12 & 12.0$\pm$0.5 & 3.0$\pm$0.2 & 1.10$\pm$0.09 & 0.22$\pm$0.15 \\	
MACSJ0329-02 & 0.450 & 0.86$\pm$0.11 & 3.69$\pm$0.21 &  & 1.73$\pm$0.19 & 15.0$\pm$0.6 & 2.0$\pm$0.2 & 0.79$\pm$0.06 & 0.21$\pm$0.04 \\	
RXJ1347-1145 & 0.451 & 1.35$\pm$0.19 & 3.68$\pm$0.85 &  & 0.71$\pm$0.10 & 14.0$\pm$0.6 & 3.0$\pm$0.2 & 1.30$\pm$0.10 & 0.35$\pm$0.13 \\	
MACSJ1311-03 & 0.494 & 0.53$\pm$0.04 & 4.49$\pm$1.05 &  & 1.00$\pm$0.13 & 9.0$\pm$0.4 & 1.0$\pm$0.1 & 0.23$\pm$0.02 & 0.05$\pm$0.12 \\	
MACSJ1149+22 & 0.544 & 5.10$\pm$1.90 & 3.05$\pm$0.58 &  & 0.53$\pm$0.10 & 19.0$\pm$0.8 & 2.0$\pm$0.2 & 1.10$\pm$0.09 & 0.36$\pm$0.11 \\	
MACSJ1423+24 & 0.545 & 0.65$\pm$0.11 & 4.95$\pm$0.13 &  & 2.34$\pm$0.45 & 11.0$\pm$0.4 & 3.0$\pm$0.2 & 0.95$\pm$0.08 & 0.19$\pm$0.02 \\	
MACSJ2129-07 & 0.570 & 3.50$\pm$3.10 & 2.44$\pm$0.17 &  & 0.69$\pm$0.21 & 18.0$\pm$0.7 & 4.0$\pm$0.3 & 1.74$\pm$0.14 & 0.72$\pm$0.06 \\	
MACSJ0647+70 & 0.584 & 6.80$\pm$1.40 & 5.13$\pm$1.79 &  & 1.00$\pm$0.26 & 11.0$\pm$0.4 & 4.0$\pm$0.3 & 10.43$\pm$0.83 & 2.03$\pm$0.26 \\	
MACSJ0744+39 & 0.686 & 0.79$\pm$0.04 & 8.20$\pm$0.43 &  & 1.18$\pm$0.16 & 5.0$\pm$0.2 & 2.0$\pm$0.2 & 0.65$\pm$0.05 & 0.08$\pm$0.03 \\	
CLJ1226+3332 & 0.890 & 1.72$\pm$0.11 & 15.00$\pm$1.99 &  & 1.47$\pm$0.09 & 15.0$\pm$0.6 & 8.0$\pm$0.6 & 6.71$\pm$0.54 & 0.45$\pm$0.08 \\	

\hline

\end{tabular}
\end{center}
\end{table} 
\end{landscape}
\pagestyle{plain}

\label{lastpage}

\end{document}